\documentclass{sig-alternate-05-2015}

\usepackage{epsfig, amsmath}
\usepackage{algorithm}
\usepackage{algpseudocode}
\usepackage{tabularx}

\usepackage[us,12hr]{datetime}
\usepackage{multirow}
\usepackage{pifont}
\usepackage{amssymb}
\DeclareMathOperator{\sgn}{sgn}
\usepackage{microtype}
\usepackage{hyperref}
\usepackage{subcaption}
\usepackage{setspace}
\usepackage{enumitem}

\newcommand{\boldpara}[1] {\vspace*{0.075in}\noindent\textbf{#1:}}
\newcommand{\boldparab}[1] {\noindent\textbf{#1:}}

\begin{document}

\title{Practical Black-Box Attacks against Machine Learning}

\numberofauthors{6}
\author{
	\alignauthor
	Nicolas Papernot\\
	\affaddr{Pennsylvania State University}\\
	\affaddr{ngp5056@cse.psu.edu}
	\alignauthor
	Patrick McDaniel\\
	\affaddr{Pennsylvania State University}\\
	\affaddr{mcdaniel@cse.psu.edu}
	\alignauthor Ian Goodfellow\titlenote{Work done while the author was at Google.}\\
	\affaddr{OpenAI}\\
	\affaddr{ian@openai.com}
	\and  
	\alignauthor Somesh Jha\\
	\affaddr{University of Wisconsin}\\
	\affaddr{jha@cs.wisc.edu}
	\alignauthor Z. Berkay Celik\\
	\affaddr{Pennsylvania State University}\\
	\affaddr{zbc102@cse.psu.edu}
	\alignauthor Ananthram Swami\\
	\affaddr{US Army Research Laboratory}\\
	\affaddr{ananthram.swami.civ@mail.mil}
}

\setcopyright{licensedusgovmixed}
\conferenceinfo{ASIA CCS '17,}{April 02 - 06, 2017, Abu Dhabi, United Arab Emirates}
\isbn{978-1-4503-4944-4/17/04}\acmPrice{\$15.00}
\doi{http://dx.doi.org/10.1145/3052973.3053009}

\maketitle

\begin{abstract}
Machine learning (ML) models, e.g., deep neural networks (DNNs), are vulnerable
to adversarial examples:  malicious inputs modified to yield erroneous model
outputs, while appearing unmodified to human observers. Potential attacks
include having malicious content like malware identified as legitimate or
controlling vehicle behavior. Yet, all existing adversarial example attacks
require knowledge of either the model internals or its training data. We
introduce the first practical demonstration of an attacker controlling a
remotely hosted DNN with no such knowledge. Indeed, the only capability of our
black-box adversary is to observe labels given by the DNN to chosen inputs. Our
attack strategy consists in training a local model to substitute for the target
DNN, using inputs synthetically generated by an adversary and labeled by the
target DNN. We use the local substitute to craft adversarial examples, and find
that they are misclassified by the targeted DNN. To perform a real-world and
properly-blinded evaluation, we attack a DNN hosted by MetaMind, an online deep
learning API. We find that their DNN misclassifies 84.24\% of the adversarial
examples crafted with our  substitute. We demonstrate the general applicability
of our strategy to many ML techniques by conducting the same attack against
models hosted by Amazon and Google, using logistic regression substitutes. They
yield adversarial examples misclassified by Amazon and Google at rates of
96.19\% and 88.94\%. We also find that this black-box attack strategy is capable
of evading defense strategies previously found to make adversarial example
crafting harder.
\end{abstract}


\section{Introduction}

A \emph{classifier} is a ML model that learns a mapping between inputs and a  set of \emph{classes}. For instance, a malware detector is a classifier taking executables as inputs and assigning them to the benign or malware class. Efforts in the security~\cite{huang2011adversarial,biggio2013evasion,papernot2015limitations,xuautomatically} and machine learning~\cite{szegedy2013intriguing,goodfellow2014explaining} communities exposed the vulnerability of classifiers to integrity
attacks.
Such attacks are often instantiated by \emph{adversarial examples}: legitimate inputs altered by adding small,
often imperceptible,
perturbations to force a learned classifier to misclassify the
resulting adversarial inputs, while remaining correctly classified by a human observer.
To illustrate, consider the following images,
potentially consumed by an autonomous vehicle~\cite{Stallkamp2012}:

\vspace{-1em}
\begin{center}
\includegraphics[width=0.75in]{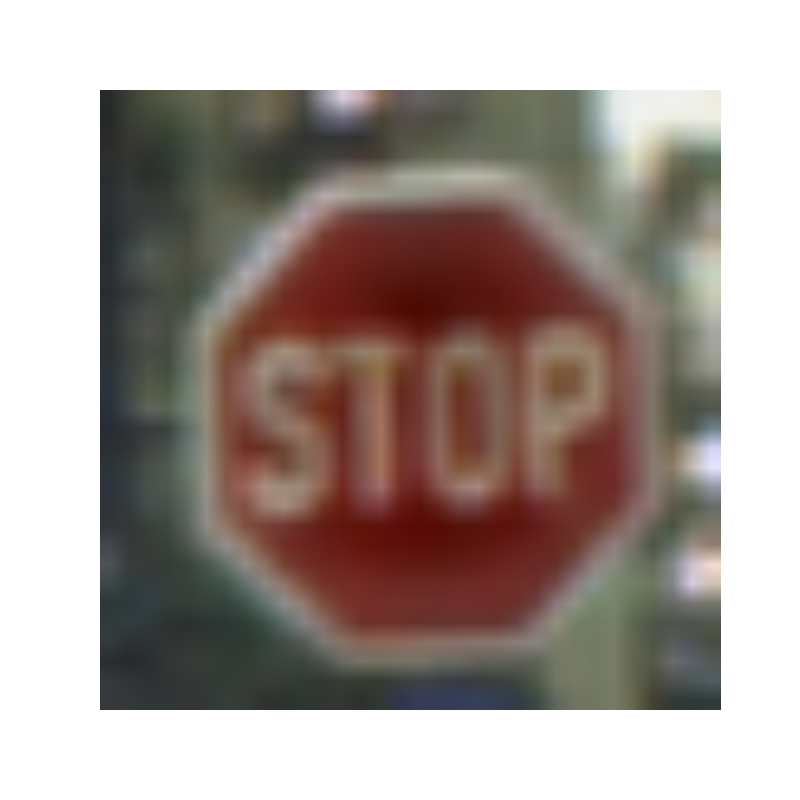} \hspace{0.2in}
\includegraphics[width=0.75in]{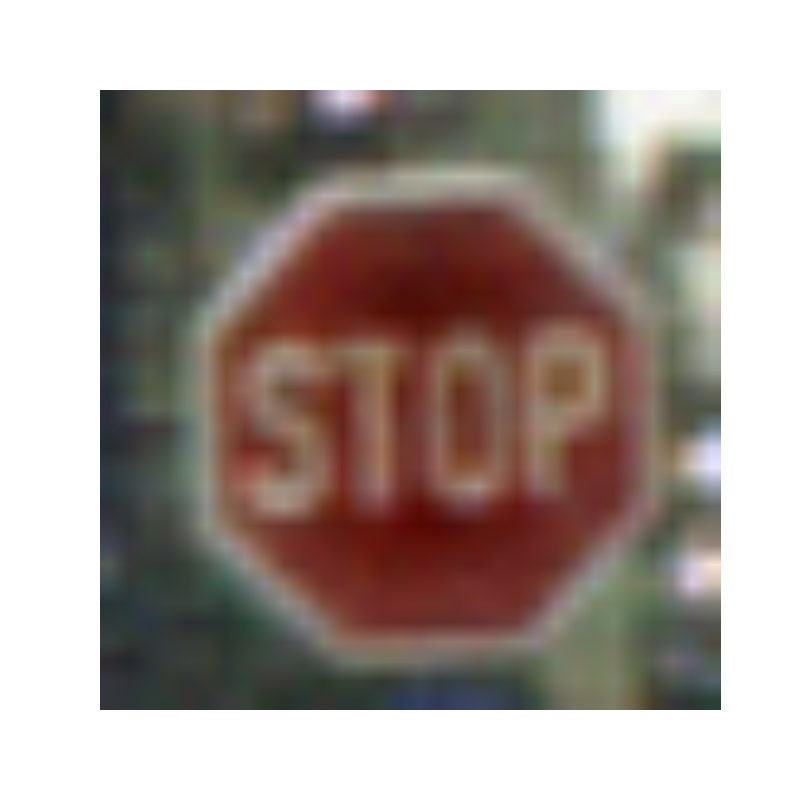}
\end{center}
\vspace{-1em}
\noindent
To humans, these images appear to be the same: our biological classifiers
(vision) identify each image as a stop sign.
The image on the left~\cite{Stallkamp2012} is indeed an ordinary image of
a stop sign.
We produced the image on the right by adding a precise perturbation that
forces a particular DNN to classify it as a yield sign, as described in Section~\ref{ssec:traffic-signs}.
Here, an adversary could potentially use the altered image to cause a car without failsafes to behave
dangerously.
This attack would require modifying
the image used internally by the car through transformations of the physical traffic sign. Related works showed the feasibility of such physical transformations for a state-of-the-art vision classifier~\cite{kurakin2016adversarial} and face recognition model~\cite{sharif2016accessorize}. 
It is thus conceivable that physical adversarial traffic signs could be generated by maliciously
modifying the sign itself, e.g., with stickers or paint. 

In this paper, we introduce the first demonstration that \emph{black-box attacks} against DNN classifiers are practical for real-world adversaries with \emph{no} knowledge about the model. We assume the adversary (a) has no information about the structure or parameters of the DNN, and (b) does not have access to any large training dataset.
The adversary's only capability is to observe labels assigned by the DNN for
chosen inputs, in a manner analog to a cryptographic oracle.

Our novel attack strategy is to train a local substitute DNN with a \emph{synthetic}
dataset: the inputs are synthetic and generated
 by the adversary, while the outputs are labels assigned by the 
target DNN and observed by the adversary. Adversarial
examples are crafted using the substitute  parameters, which are known to
us.
They are not only misclassified by the substitute  but
also by the target DNN, because both models have similar decision boundaries.

This is a considerable departure from previous work, which evaluated
perturbations required to craft adversarial examples using either:
(a) detailed knowledge of the DNN architecture and
parameters~\cite{biggio2013evasion,goodfellow2014explaining,papernot2015limitations,szegedy2013intriguing}, 
or (b) an independently collected training set to fit an auxiliary model~\cite{biggio2013evasion,goodfellow2014explaining,szegedy2013intriguing}.
This limited their applicability to strong adversaries
capable of
gaining insider knowledge of the targeted ML model, or
collecting large labeled training sets. We release assumption (a) by learning a substitute: it gives us the benefit of having full access to the model and apply previous adversarial example crafting methods. 
We release assumption (b) by replacing the independently collected training set with a synthetic dataset constructed by the adversary with synthetic inputs and labeled by observing the target DNN's output. 

Our threat model thus corresponds to the real-world scenario of users interacting with
classifiers hosted remotely by a third-party keeping the model
internals  secret. In fact, we instantiate our attack
against  classifiers automatically trained by MetaMind, Amazon, and
Google.
We are
 able to access them only after training is completed.
Thus, we provide the first correctly blinded experiments concerning
adversarial examples as a security risk.

We show that our black-box attack is applicable to many remote systems taking decisions based on ML, because it combines three key properties: (a) the capabilities required are limited to observing output class labels, (b) the number of labels queried is limited, and (c) the approach applies and scales to different ML classifier types (see Section~\ref{sec:generalization}), in addition to state-of-the-art DNNs. In contrast, previous work failed to simultaneously provide all of these three key
properties~\cite{goodfellow2014explaining,szegedy2013intriguing,laskov2014practical,tramer2016stealing,xuautomatically}. 
Our contributions are:
\vspace*{-0.05in}
\begin{itemize}
\setlength\itemsep{-0cm}
\item We introduce in Section~\ref{sec:methodology} an attack 
against black-box DNN classifiers. It crafts adversarial examples without knowledge of the classifier training data or model. To do so, a synthetic dataset is constructed by the adversary to train a substitute for the targeted DNN classifier.
\item In Section~\ref{sec:validation}, we instantiate the attack against a remote DNN classifier hosted by MetaMind. The DNN misclassifies $84.24\%$
of the adversarial inputs crafted. 
\item The attack is calibrated in
Section~\ref{sec:evaluation} to (a) reduce the number of queries made to the target model and (b)
maximize
misclassification of adversarial examples. 
\item We generalize the attack to other ML classifiers like logistic regression. In Section~\ref{sec:generalization}, we target models hosted by Amazon and Google. They  misclassify adversarial examples at rates of $96.19\%$ and $88.94\%$. 
\item Section~\ref{sec:defenses} shows that our attack evades defenses proposed in the literature because the substitute trained by the adversary is unaffected by defenses deployed on the targeted oracle model to reduce its vulnerability.
\item In Appendix B, we provide an intuition of why adversarial examples crafted with the substitute also mislead target models by empirically observing that substitutes have 
gradients correlated to the target's. 
\end{itemize}
\vspace*{-0.1in}
 \boldpara{Disclosure} We disclosed our attacks to MetaMind, Amazon, and Google. Note that no damage was caused as we demonstrated control of models created for our own account.


\section{About Deep Neural Networks}

We provide preliminaries of deep learning to enable
understanding of our threat model and attack. We refer readers
interested to the more detailed presentation in~\cite{Goodfellow-et-al-2016-Book}.

\begin{figure}[t]
	\centering
	\includegraphics[width=0.9\columnwidth]{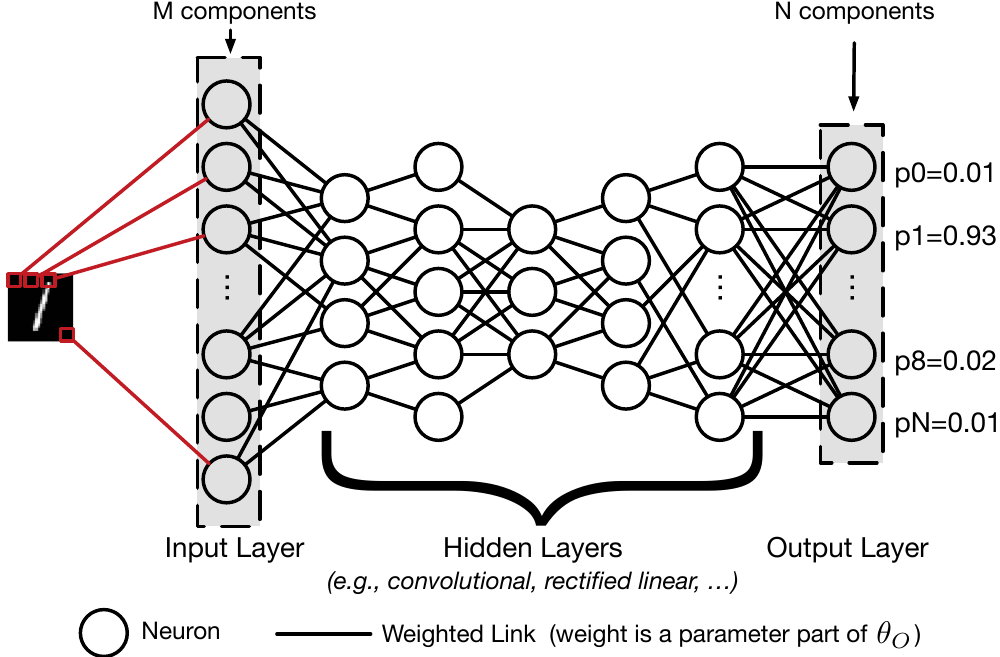}
	\caption{\textbf{DNN Classifier:} the model processes an image of a handwritten digit and outputs the probility of it being in one of the $N=10$ classes for digits 0 to 9 (from~\cite{papernot2015distillation}).}
	\label{fig:classifier-dnn}
\end{figure}

A \emph{deep neural network} (DNN), as illustrated in
Figure~\ref{fig:classifier-dnn}, is a ML technique that uses a
hierarchical composition of $n$ parametric functions to model an input
$\vec{x}$. Each function
$f_i$ for $i\in 1..n$ is modeled using a layer of neurons, which are
 elementary computing units applying an \emph{activation function}
to the previous layer's weighted representation of the input to generate a new
representation. Each layer is parameterized by a weight vector
$\theta_i$ (we omit the vector notation) impacting each neuron's activation. Such weights hold
the knowledge of a DNN model $F$ and are evaluated during its
training phase, as detailed below. Thus, a DNN defines and computes: 
\begin{equation}
F(\vec{x})=f_n \left( \theta_n, f_{n-1}\left( \theta_{n-1}, \text{  ... } f_2\left(\theta_2, f_1 \left(\theta_1, \vec{x}\right)\right)\right)\right)
\end{equation}

The \emph{training phase} of a DNN $F$ learns values for its 
parameters $\theta_F=\{\theta_1, ..., \theta_n\}$. We focus on  classification tasks, where the
goal is to assign inputs a label among a predefined set of labels. The DNN is given a large set of known input-output pairs
$(\vec{x},\vec{y})$ and it adjusts weight parameters to reduce a cost
quantifying the prediction error between the prediction $F(\vec{x})$ and
the correct output $\vec{y}$. The adjustment is typically performed using
techniques derived from the backpropagation
algorithm. Briefly, such techniques
successively propagate error gradients with respect to network parameters from
the network's output layer to its input layer.

During the \emph{test phase}, the DNN is deployed with a fixed set of
parameters $\theta_F$ to make predictions on inputs unseen during
training. We consider classifiers: the DNN produces a probability vector
$F(\vec{x})$ encoding its belief of input $\vec{x}$ being in each of the
 classes (cf. Figure~\ref{fig:classifier-dnn}). The weight parameters
$\theta_F$ hold the model knowledge acquired by training. Ideally,
the model should
generalize and make accurate predictions for inputs outside of the domain
explored during training. However, attacks manipulating DNN inputs with adversarial
examples showed this is not the case in practice~\cite{goodfellow2014explaining,papernot2015limitations,szegedy2013intriguing}.


\section{Threat Model}
\label{ssec:threat-model}

A taxonomy of adversaries against DNN classifiers is found in~\cite{papernot2015limitations}. 
In our work, the adversary
seeks to force a classifier to misclassify inputs in any class different from
their correct class. To achieve this, we consider a weak adversary 
with access to the DNN output only. The adversary has no knowledge of the architectural
choices made to design the DNN, which include the number, type, and size of
layers, nor of the training data used to learn the DNN's parameters. Such attacks are referred to as {\it black box}, where adversaries need not know internal details of a 
system to compromise it.

\boldparab{Targeted Model} We
consider attackers targeting a multi-class DNN classifier. It
outputs probability vectors, where each vector component
encodes the DNN's belief of the input being part of one of the predefined
classes. We consider the ongoing example of
a DNN classifying images, as shown in
Figure~\ref{fig:classifier-dnn}. Such DNNs can be used to classify
handwritten digits into classes associated with digits from 0 to 9, images
of objects in a fixed number of categories, or images of traffic signs into
classes identifying its type (STOP, yield, ...).

\boldpara{Adversarial Capabilities} The \emph{oracle} $O$ is the targeted DNN. Its name refers to the only capability of the adversary:
accessing the label $\tilde{O}(\vec{x})$ for
any input $\vec{x}$ by querying oracle $O$. The
output label $\tilde{O}(\vec{x})$ is 
the index of the class assigned the largest probability by the DNN:
\begin{equation}
\tilde{O}(\vec{x}) = \arg\max_{j\in0..N-1} O_j(\vec{x})
\end{equation}
where $O_j(\vec{x})$ is the $j$-th component of the probability vector
$O(\vec{x})$ output by DNN $O$. Distinguishing between
labels and probabilities makes adversaries 
realistic (they more often have access to labels than probabilities) but weaker: labels encode less information
about the model's learned behavior. \textbf{Accessing
labels $\tilde{O}$ produced by the DNN $O$ is the only capability assumed in our
threat model.}
We do not have
access to the oracle internals or training data.

\begin{figure}[t]
		\begin{center}
	\includegraphics[width=0.85\columnwidth]{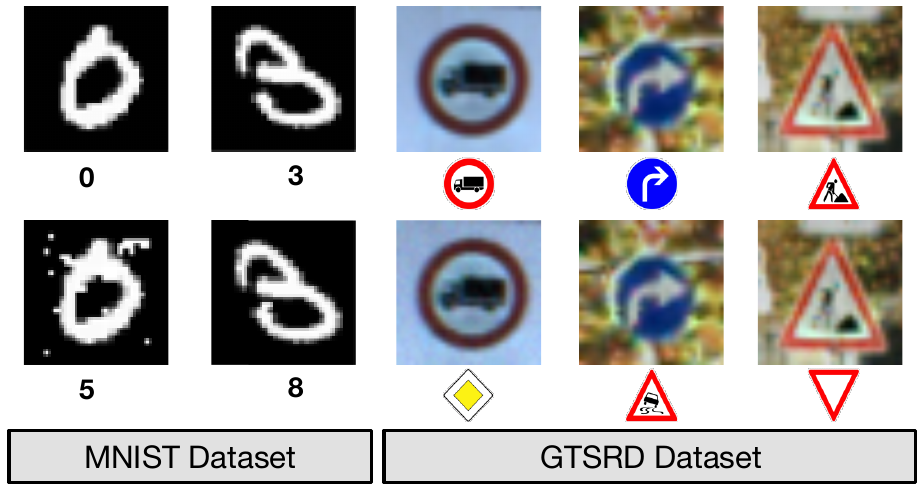}
		\end{center}
	\caption{\textbf{Adversarial samples} (misclassified) in the bottom row are created from the legitimate samples~\cite{lecun1998mnist,Stallkamp2012} in the top row. The DNN outputs are identified below the samples.}
	\label{fig:adversarial-samples}
	\vspace*{-0.09in}
\end{figure}

\boldpara{Adversarial Goal} We want to produce a minimally altered version of any input $\vec{x}$, named
\emph{adversarial sample}, and denoted $\vec{x^*}$, misclassified by oracle $O$: $\tilde{O}(\vec{x^*}) \neq \tilde{O}(\vec{x})$. This corresponds to an attack on the
oracle's output integrity. Adversarial samples solve the following
optimization problem:
\begin{equation}
\label{eq:adv-sample}
\vec{x^*}= \vec{x} + \arg\min \{\vec{z} :  \tilde{O}(\vec{x}+\vec{z}) \neq \tilde{O}(\vec{x}) \}  =  \vec{x} + \delta_{\vec{x}} 
\end{equation}
Examples of adversarial samples can be found in
Figure~\ref{fig:adversarial-samples}. The first row contains legitimate samples
and the second corresponding adversarial samples that are
misclassified. This misclassification must be achieved by adding a minimal
perturbation $\delta\vec{x}$ so as to evade human detection. Even with total
knowledge of the architecture used to train model $O$ and its
parameters resulting from training, finding such a minimal perturbation is not
trivial, as properties of DNNs preclude the optimization
problem from being linear or convex. This is exacerbated by our threat
model: removing knowledge of model $O$'s architecture and training data makes it harder to  find a perturbation such that $\tilde{O}(\vec{x}+\delta\vec{x})
\neq \tilde{O}(\vec{x})$ holds.

\vspace*{0.05in}

In Appendix C, we give a presentation of attacks conducted in related threat models---with stronger assumptions.


\section{Black-Box Attack Strategy}
\label{sec:methodology}

We introduce our black-box attack.
As stated in Section~\ref{ssec:threat-model}, the adversary wants to craft
inputs misclassified by the ML model using the sole
capability of accessing the label $\tilde{O}(\vec{x})$ assigned by classifier for any chosen
input $\vec{x}$. The strategy is to learn a
\emph{substitute} for the target model using a synthetic dataset generated by the adversary and labeled by observing the oracle output. Then, adversarial examples are crafted using this substitute. We expect the target DNN to misclassify them due to transferability
between architectures~\cite{szegedy2013intriguing,goodfellow2014explaining}

To understand the difficulty of conducting the attack under this threat model,
recall Equation~\ref{eq:adv-sample} formalizing the adversarial goal of
finding a minimal perturbation that forces
the targeted oracle to misclassify. A closed form solution cannot be found when the target is a non-convex ML model: e.g., a DNN.
The basis for most adversarial
attacks~\cite{goodfellow2014explaining,papernot2015limitations,szegedy2013intriguing}
is to approximate its solution 
using gradient-based optimization on functions defined by a DNN.
Because evaluating these functions and their gradients requires knowledge
of the DNN architecture and parameters, such an attack is not possible
under our black-box scenario.
It was shown that adversaries with access 
to an independently collected labeled training set from the same population distribution
than the oracle could train a model with a different architecture and use it as a substitute~\cite{szegedy2013intriguing}:
adversarial examples designed to manipulate the substitute
are often misclassified by the targeted model.
However, many modern machine learning systems require large and
expensive training sets for training. For instance, 
we consider models trained with several tens of thousands 
of labeled examples. This makes attacks based
on this paradigm unfeasible for adversaries without large labeled datasets.

In this paper, we show black-box attacks can be accomplished at
a much lower cost, without labeling an independent training
set.
In our approach, to enable the adversary to train a substitute model without a real labeled dataset, we use the target DNN as an oracle to construct a synthetic dataset. The inputs are synthetically generated and the outputs are labels observed from the oracle. 
Using this synthetic dataset, 
the attacker builds an approximation $F$ of the model $O$
learned by the oracle. This \emph{substitute network} $F$ is then used to craft adversarial samples misclassified
by $F$ 
Indeed, with its full knowledge of the substitute DNN $F$ parameters, the adversary can use one of the previously
described attacks~\cite{goodfellow2014explaining,papernot2015limitations} to craft
adversarial samples misclassified by $F$. As long as the transferability
property holds between $F$ and $O$, adversarial samples crafted for $F$ will
also be misclassified by $O$. This leads us to propose the following 
 strategy: 
\vspace*{-0.07in}
\begin{enumerate}
	\item \textbf{Substitute Model Training:} the attacker queries the oracle with synthetic inputs selected by a Jacobian-based heuristic to build a model $F$ approximating the oracle model $O$'s decision boundaries. 
	\item \textbf{Adversarial Sample Crafting:} the attacker uses substitute network $F$ to craft adversarial samples, which are then misclassified by oracle $O$ due to the transferability of adversarial samples. 
\end{enumerate} 
\vspace*{-0.07in}
\noindent 

\subsection{Substitute Model Training}

\begin{figure*}[t]
	\begin{center}
		\includegraphics[width=0.91\textwidth]{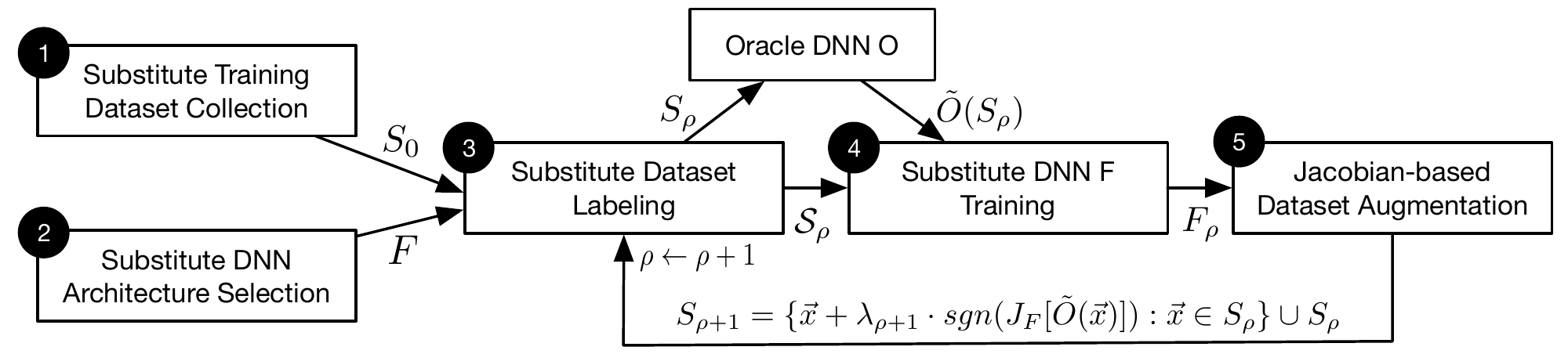}
		\caption{\textbf{Training of the substitute DNN $F$}: the attacker (1) collects an initial substitute training set $S_0$ and (2) selects an architecture $F$. Using oracle $\tilde{O}$, the attacker (3) labels $S_0$ and (4) trains substitute $F$. After (5) Jacobian-based dataset augmentation, steps (3) through (5) are repeated for several substitute epochs $\rho$.}
		\label{fig:substitute-training}
	\end{center}
\end{figure*}

Training a substitute model $F$ approximating oracle $O$ is challenging because  we must: (1) select an architecture for our substitute without knowledge of the
targeted oracle's architecture, and (2) limit the number of queries
made to the oracle in order to ensure that the approach is tractable. Our
approach, illustrated in Figure~\ref{fig:substitute-training}, overcomes these challenges mainly by introducing a synthetic data generation technique, the \emph{Jacobian-based Dataset
Augmentation}. We emphasize that \emph{this technique is not
designed to maximize the substitute DNN's accuracy but rather ensure that it
approximates the oracle's decision boundaries with few label queries}. 

\boldpara{Substitute Architecture} This factor is not the most
limiting as the adversary must at least have some partial knowledge of the
oracle input  (e.g., images, text) and 
expected output (e.g., classification). The adversary can thus use
an architecture adapted to the input-output relation. For instance, a
convolutional neural network is suitable for image classification. Furthermore,
we show in Section~\ref{sec:evaluation} that the type, number, and size of layers used
in the substitute DNN have relatively little
impact on the success of the attack. Adversaries can also consider performing
an architecture exploration and train several substitute models before
selecting the one yielding the highest attack success.

\boldpara{Generating a Synthetic Dataset} To better understand
the need for synthetic data, note that we could potentially make an infinite number of queries
to obtain the oracle's output $O(\vec{x})$ for any input $\vec{x}$ belonging to
the input domain. This would provide us with a copy of the oracle.
However, this is simply not tractable: consider a DNN with $M$ input
components, each taking discrete values among a set of $K$ possible values, the
number of possible inputs to be queried is $K^M$. The intractability is even
more apparent for inputs in the continuous domain. Furthermore, making a large
number of queries renders the adversarial behavior easy to detect.

A natural alternative is to resort to randomly selecting additional 
points to be queried. For instance, we tried using Gaussian noise to select
points on which to train substitutes. However, the resulting models 
were not able to learn by querying the oracle. This is likely due to
noise not being representative of the input distribution. 
To address this issue, we thus introduce a heuristic efficiently 
exploring the input domain and, as shown in
Sections~\ref{sec:validation} and~\ref{sec:evaluation}, drastically limits the
number of oracle queries. Furthermore, our technique also ensures that the
substitute DNN is an approximation of the targeted DNN i.e. it
learns similar decision boundaries.

The heuristic used to generate synthetic training inputs is based on identifying directions in which the model's output is
varying, around an initial set of training points. Such directions 
intuitively require more input-output pairs
 to capture the output variations of the target DNN $O$. Therefore, to get a
substitute DNN accurately approximating the oracle's decision boundaries,
the heuristic prioritizes these samples when querying the oracle for labels. These directions are identified with the substitute DNN's Jacobian matrix $J_F$, which is
evaluated at several input points $\vec{x}$ (how these points are chosen is
described below). Precisely, the adversary evaluates the sign of the Jacobian matrix dimension corresponding to the label assigned to input $\vec{x}$ by the oracle:
$\sgn\left(J_{F}(\vec{x})[\tilde{O}(\vec{x})]\right)
$. To obtain a new synthetic training point, a term $
\lambda \cdot \sgn\left(J_{F}(\vec{x})[\tilde{O}(\vec{x})]\right)
$ is added to the original point $\vec{x}$. We name this technique \emph{Jacobian-based
Dataset Augmentation}. We base our substitute training algorithm on the idea of
iteratively refining the model in directions identified using the Jacobian.

\begin{algorithm}[t]
\caption{\textbf{- Substitute DNN Training:} for oracle $\tilde{O}$, a maximum number $max_{\rho}$ of substitute training epochs, a substitute architecture $F$, and an initial training set $S_0$.}
\label{alg:substitute-training}
\begin{algorithmic}[1]
	\Require $\tilde{O}$, $max_{\rho}$, $S_0$, $\lambda$
	\State Define architecture $F$
	\For{$\rho \in 0 \mbox{ .. } max_{\rho}-1$}
		\State // {\it Label the substitute training set}
		\State $D \leftarrow \left\{ (\vec{x}, \tilde{O}(\vec{x})) : \vec{x}\in S_\rho \right\}$
		\State // {\it Train $F$ on $D$ to evaluate parameters $\theta_F$}
		\State $\theta_F \leftarrow \mbox{train}(F,D)$
		\State // {\it Perform Jacobian-based dataset augmentation}
		\State $S_{\rho+1} \leftarrow \{ \vec{x} + \lambda\cdot \sgn(J_{F}[\tilde{O}(\vec{x})]) : \vec{x} \in S_\rho \} \cup S_\rho$
	\EndFor
	\State \Return $\theta_F$
\end{algorithmic}
\end {algorithm}

\boldpara{Substitute DNN Training Algorithm} We now describe the five-step training procedure outlined in Algorithm~\ref{alg:substitute-training}:
\begin{itemize}
	\item \textbf{Initial Collection (1):} The adversary collects a very small set $S_0$ of inputs representative of the input domain. For instance, if the targeted oracle $O$ classifies handwritten digits, the adversary collects 10 images of each digit 0 through 9. We show in Section~\ref{sec:validation} that this set does not necessarily have to come from the distribution from which the targeted oracle was trained.
	\item \textbf{Architecture Selection (2):} The adversary selects an architecture to be trained as the substitute $F$. Again, this can be done using high-level knowledge of the classification task performed by the oracle (e.g., convolutional networks are appropriate for vision) 
	\item \textbf{Substitute Training:} The adversary iteratively trains more
	accurate substitute DNNs $F_\rho$ by repeating the following for $\rho \in 0.. \rho_{max}$:
	\begin{itemize}
	\item \textbf{Labeling (3):} By querying for the labels $\tilde{O}(\vec{x})$ output by oracle $O$, the adversary labels each sample $\vec{x}\in S_\rho$ in its initial substitute training set $S_\rho$.
	\item \textbf{Training (4):} The adversary trains the architecture chosen at step (2) using substitute training set $S_\rho$ in conjunction with classical training techniques.
	\item \textbf{Augmentation (5):} The adversary applies our augmentation technique on the initial substitute training set $S_\rho$ to produce a larger substitute training set $S_{\rho+1}$ with more synthetic training points. This new training set better represents the model's decision boundaries. The adversary repeats steps (3) and (4) with the augmented set $S_{\rho+1}$.  
	\end{itemize}
\end{itemize}
Step (3) is repeated several times to increase the substitute DNN's accuracy and the similarity of its decision boundaries with the oracle. We introduce the term \emph{substitute training epoch}, indexed with $\rho$, to refer to each iteration performed. This leads to this formalization of the Jacobian-based Dataset Augmentation performed at step (5) of our substitute training algorithm to find more synthetic training points:
\begin{equation}
\label{eq:jacobian-augmentation}
S_{\rho+1}= \{ \vec{x} + \lambda\cdot \sgn(J_{F}[\tilde{O}(\vec{x})]) : \vec{x} \in S_\rho \} \cup S_\rho
\end{equation} 
where $\lambda$ is a parameter of the augmentation: it defines the size of the step taken in the sensitive direction identified by the Jacobian matrix to augment the set $S_\rho$ into $S_{\rho+1}$.

\subsection{Adversarial Sample Crafting}

Once the adversary trained a substitute DNN, it uses it to craft adversarial samples. This is performed by implementing two
previously introduced approaches described in~\cite{goodfellow2014explaining,
papernot2015limitations}. We  provide an overview of the two approaches, namely the \emph{Goodfellow et al. algorithm} and the
\emph{Papernot et al. algorithm}. Both techniques share a similar intuition of
evaluating the model's sensitivity to input modifications in order to select a
small perturbation achieving the misclassification goal\footnote{Our attack can be implemented with other adversarial example algorithms. We focus on these two in our evaluation.}.

\boldpara{Goodfellow et al. algorithm} This algorithm is also known as the
\emph{fast gradient sign method}~\cite{goodfellow2014explaining}.
Given a model $F$ with an associated
cost function $c(F,\vec{x},y)$, the adversary crafts an adversarial sample
$\vec{x^*}=\vec{x}+\delta_{\vec{x}}$ for a given legitimate sample $\vec{x}$ by
computing the following perturbation:
\begin{equation}
\label{eq:goodfellow-perturbation}
\delta_{\vec{x}} = \varepsilon \sgn(\nabla_{\vec{x}} c(F, \vec{x}, y))
\end{equation}
where perturbation $\sgn(\nabla_{\vec{x}} c(F, \vec{x}, y))$ is the sign of the
model's cost function
\footnote{As described here, the method causes simple misclassification.
It has been extended to achieve chosen target classes.
}
gradient. 
The cost gradient is computed with respect to
$\vec{x}$ using sample $\vec{x}$ and label $y$ as inputs. The value of the
\emph{input variation parameter} $\varepsilon$ factoring the sign matrix controls the
perturbation's amplitude. Increasing its value increases the likelihood of
$\vec{x^*}$ being misclassified by model $F$ but on the contrary makes
adversarial samples easier to detect by humans. In
Section~\ref{sec:evaluation}, we evaluate the impact of parameter $\varepsilon$
on the successfulness of our attack. 

\boldpara{Papernot et al. algorithm} This algorithm is suitable for
source-target misclassification attacks where adversaries seek to take samples
from any legitimate source class to any chosen target
class~\cite{papernot2015limitations}. Misclassification attacks are a special
case of source-target misclassifications, where the target class can
be any class different from the legitimate source class. Given model $F$, the
adversary crafts an adversarial sample $\vec{x^*} = \vec{x} + \delta_{\vec{x}}$
for a given legitimate sample $\vec{x}$ by adding a perturbation
$\delta_{\vec{x}}$ to a subset of the input components $\vec{x}_i$.

To choose input components forming perturbation $\delta_{\vec{x}}$,
components are sorted by decreasing adversarial saliency value. The
adversarial saliency value $S(\vec{x},t)[i]$ of component $i$ for an
adversarial target class $t$ is defined as:
\begin{equation}
\label{eq:saliency-map-increasing-features}
S(\vec{x},t)[i] = \left\lbrace
\begin{array}{c}
0  \mbox{ if }   \frac{\partial F_t}{\partial \vec{x}_i}(\vec{x})<0  \mbox{ or } \sum_{j\neq t} \frac{\partial F_j}{\partial \vec{x}_i}(\vec{x})>0\\
 \frac{\partial F_t}{\partial \vec{x}_i}(\vec{x})  \left| \sum_{j\neq t}  \frac{\partial F_j}{\partial \vec{x}_i}(\vec{x})\right| \mbox{ otherwise}
\end{array}\right.
\end{equation}
where matrix $J_F=\left[\frac{\partial F_j}{\partial \vec{x}_i}\right]_{ij}$ is
the model's Jacobian matrix. Input components $i$ are added to perturbation
$\delta_{\vec{x}}$ in order of decreasing adversarial saliency value
$S(\vec{x},t)[i]$  until the resulting adversarial sample $\vec{x^*} = \vec{x}
+ \delta_{\vec{x}}$ is misclassified by $F$. The perturbation introduced
for each selected input component can vary: greater perturbation 
reduce the number of components perturbed to achieve misclassification. 

Each algorithm has its benefits and drawbacks. The Goodfellow algorithm is
well suited for fast crafting of many adversarial samples
with relatively large perturbations thus potentially easier to detect.
The Papernot algorithm reduces
perturbations at the expense of a greater computing cost.


\section{Validation of the Attack}
\label{sec:validation}

We validate our attack against remote and local classifiers. We first apply it to target a DNN remotely
provided by MetaMind, through their API\footnote{\label{metamind-footnote}The API can be
accessed online at www.metamind.io} that allows a user to train classifiers using
deep learning. The API returns labels produced by the DNN for any given input
but does not provide access to the DNN. This corresponds to the
oracle described in our threat model. We show that: 
\vspace*{-0.03in}
\begin{itemize}
	\item An adversary
	using our attack can reliably force the DNN trained
	using MetaMind on MNIST~\cite{lecun1998mnist} to misclassify
	$84.24\%$ of adversarial examples crafted with a perturbation not affecting human recognition.
	\item A second oracle trained locally with the German Traffic
	Signs Recognition Benchmark (GTSRB)~\cite{Stallkamp2012}, 
	can be
	forced to misclassify more than $64.24\%$ of altered inputs  without affecting human recognition.
\end{itemize}

\subsection{Attack against the MetaMind Oracle}

\boldpara{Description of the Oracle} We used the MNIST handwritten digit dataset to train the DNN~\cite{lecun1998mnist}. It comprises $60,000$ training and $10,000$ test images of handwritten digits. The task associated with the dataset is 
 to identify the digit corresponding to each image. Each $28$x$28$
grayscale sample is encoded as a vector of pixel intensities in the
interval $[0,1]$ and obtained by reading the image pixel matrix row-wise.

We registered for an API key on MetaMind's website, which gave us access to
three functionalities: dataset upload, automated model training, and model
prediction querying. We uploaded the $50,000$ samples included in the MNIST
training set to MetaMind and then used the API to train a classifier
on the dataset. We emphasize that training is automated: we have no
access to the training algorithm, model architecture, or model parameters. All
we are given is the accuracy of the resulting model, computed by MetaMind using
a validation set created by isolating $10\%$ of the training samples. Details can be found on MetaMind's
website.

Training took 36 hours to return a classifier
with a $94.97\%$ accuracy. This  performance cannot be
improved as we cannot access or modify the model's specifications and
training algorithm. Once training is completed, we could access the model
predictions, for any input of our choice, through the API. Predictions take the
form of a class label.
This
corresponds to the threat model described in
Section~\ref{ssec:threat-model}.

\boldpara{Initial Substitute Training Sets} First, the adversary collects an initial substitute training
set. We describe two such sets used to attack the MetaMind
oracle:

\begin{itemize}
\item MNIST subset: This initial substitute training
set is made of $150$ samples from the MNIST test set. 
They differ from those used by the oracle for training as test and training
sets are distinct. We assume adversaries can collect such a limited sample set 
under the threat model described in Section~\ref{ssec:threat-model} with
minimal knowledge of the oracle task: here, handwritten 
digit classification. 
\item Handcrafted set: To ensure our results
do not stem from similarities between the MNIST test and training sets, we
also consider a \emph{handcrafted} initial substitute training set. We handcrafted $100$ samples by handwriting $10$ digits for each
class between $0$ and $9$ with a
laptop trackpad. We then adapted them to the MNIST format of $28$x$28$
grayscale pixels. Some are shown below. 
\end{itemize}
\begin{center}
\includegraphics[width=\columnwidth]{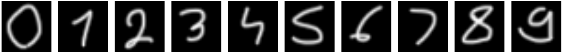}
\end{center}

\boldpara{Substitute DNN Training} The adversary uses the initial substitute training sets and the oracle to train 
subsitute DNNs. Our substitute architecture A, a standard for
image classification, is described
in Table~\ref{table:dnn-architectures} (cf. appendix). 
The substitute DNN is trained on our
machine for $6$ substitute epochs.
During
each of these $6$ epochs, the model is trained for $10$ epochs
from scratch with a learning rate of $10^{-2}$ and momentum of $0.9$. Between
substitute epochs, we perform a Jacobian-based dataset augmentation with a step
size of $\lambda = 0.1$ to generate additional synthetic training data, which we label using the MetaMind oracle. 

The accuracy of the two substitute DNNs is reported in
Figure~\ref{table-accuracy-metamind-substitutes}. It is computed with the MNIST
test set (minus the $150$ samples used in the first initial substitute training
set). The adversary does \emph{not} have access to this full test set: we
solely use it to analyze our results. The two substitute
DNNs respectively achieve a $81.20\%$ and $67.00\%$ accuracy on the MNIST test set after $6$ substitute training epochs. These accuracies fall
short of current state-of-the-art accuracies on this task. However, the adversary has access to a limited number of
samples (in this case $6,400=100\times 2^6$ instead of $50,000$ for
state-of-the-art models). Furthermore, the adversarial goal is to craft
adversarial samples misclassified by the oracle. \emph{Instead of learning a
substitute DNN with optimal accuracy, the adversary is interested in
learning a substitute capable of mimicking the oracle decision
boundaries}.

\boldparab{Adversarial Sample Crafting} Using the substitute DNNs, we then craft adversarial samples using Goodfellow's algorithm. We decided to use the $10,000$
samples from the MNIST test set as our legitimate samples.\footnote{Again, 
adversaries do not need access to the dataset and can use any legitimate sample
of their choice to craft adversarial samples. We use it in order to
show that expected inputs can be misclassified on a large scale.}  We evaluate
 sample crafting using two metrics: \emph{success rate} and
\emph{transferability}. The \emph{success rate} is the proportion of 
adversarial samples misclassified by the substitute DNN.
Our goal is to verify whether these samples are also misclassified by
the oracle or not. Therefore, the \emph{transferability of adversarial samples}
refers to the oracle misclassification rate of adversarial samples crafted
using the substitute DNN.

\begin{figure}[t]
	\centering
	\includegraphics[width=\columnwidth]{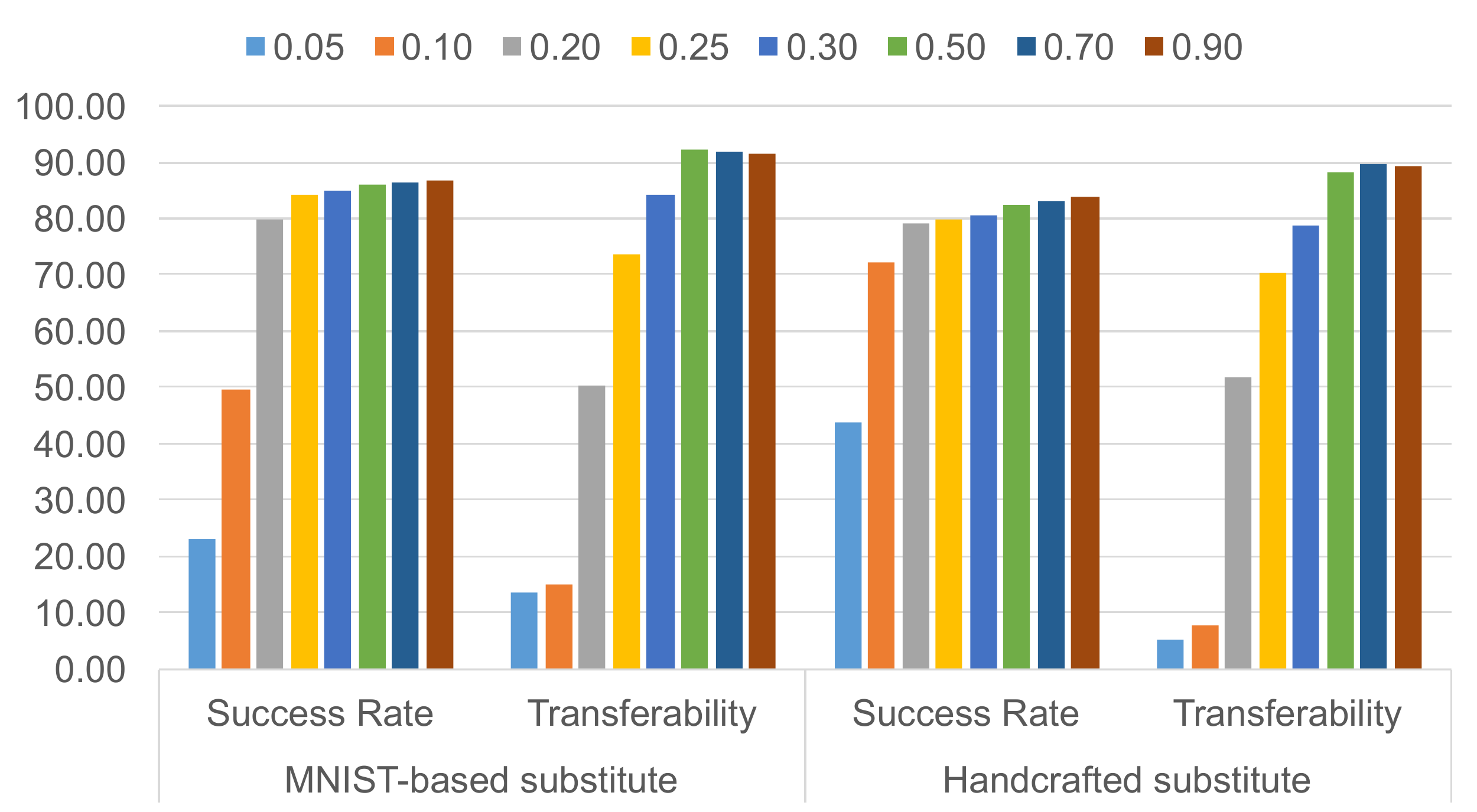}
	\caption{\textbf{Success Rate and Transferability of Adversarial Samples for the MetaMind attacks:} performed using  MNIST-based and handcrafted substitutes: each bar corresponds to a different perturbation input variation.}
	\label{fig:metamind-stats}
\end{figure}

Figure~\ref{fig:metamind-stats} details both metrics
for each substitute DNN and for several values of the input variation
$\varepsilon$ (cf. Equation~\ref{eq:goodfellow-perturbation}). Transferability reaches $84.24\%$ for the first substitute
DNN and $78.72\%$ for the second, with input variations of
$\varepsilon=0.3$. Our attack strategy is thus effectively able to severely
damage the output integrity of the MetaMind oracle. Using the 
substitute training set handcrafted by the adversary\
 limits the transferability of adversarial samples
when compared to the substitute set extracted from MNIST data,
for all input variations except $\varepsilon=0.2$.
Yet, the transferability of both substitutes is similar, corroborating that our
attack can be executed without access to any of the oracle's training data.

{

\begin{figure}[t]
	\centering
	\begin{tabular}{|c| c|c|}
		\hline
		\textbf{Substitute} & \multicolumn{2}{c|}{\textbf{Initial Substitute Training Set from}}  \\ 
		\textbf{Epoch} & \textbf{MNIST test set} & \textbf{Handcrafted digits} \\ \hline \hline
		0 & $24.86\%$ & $18.70\%$ \\ \hline 
		1 & $41.37\%$ & $19.89\%$ \\ \hline 
		2 & $65.38\%$ & $29.79\%$ \\ \hline 
		3 & $74.86\%$ & $36.87\%$ \\ \hline 
		4 & $80.36\%$ & $40.64\%$ \\ \hline 
		5 & $79.18\%$ & $56.95\%$ \\ \hline 
		6 & $81.20\%$ & $67.00\%$ \\ \hline 
	\end{tabular}
	\caption{\textbf{Substitute DNN Accuracies:} each column corresponds to an initial substitute training set:
		150 MNIST test samples, and handcrafted digits.
		Accuracy is reported on the unused 9,850 MNIST test samples.
	}
	\label{table-accuracy-metamind-substitutes}
\end{figure}

}

\begin{figure}[t]
	\centering
	\includegraphics[width=\columnwidth]{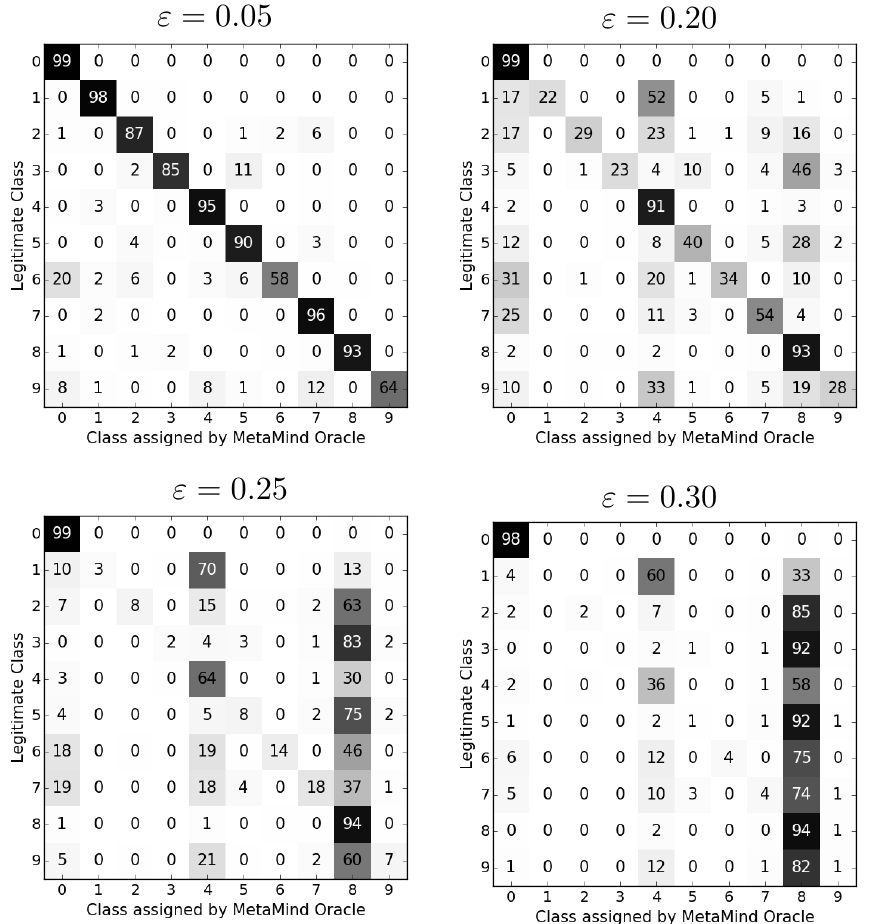}
	\caption{\textbf{MetaMind Oracle Confusion Matrices for different input variations $\varepsilon$.} Cell $(x,y)$ indicates the share of digit $y$ instances classified by the oracle as digit $x$.}
	\label{fig:metamind-confusion-matrices}
\end{figure}

To analyze the labels assigned by the MetaMind oracle, we plot confusion
matrices  for adversarial samples crafted using the first substitute DNN with
$4$ values of $\varepsilon$. In
Figure~\ref{fig:metamind-confusion-matrices}, rates on the diagonal indicate
the proportion of samples correctly classified by the oracle for each of the
$10$ classes. Off-diagonal values are the proportion of samples
misclassified in a wrong class. For instance, cell $(8,3)$ in the third matrix
indicates that $89\%$ instances of a $3$ are classified as a $8$ by the oracle
when perturbed with an input variation of $\varepsilon=0.25$.
Confusion matrices converge to most samples being classified as
$4$s and $8$s as $\varepsilon$ increases. This could be due to DNNs more easily classifying inputs in
these classes~\cite{papernot2015limitations}. 

\subsection{Attacking an oracle for the GTSRB}
\label{ssec:traffic-signs}

We now validate our attack on a different dataset, using
an oracle trained locally to recognize traffic signs on the GTSRB
dataset. The attack achieves higher transferability rates at lower distortions
compared to the MNIST oracle.

\boldpara{Oracle Description} The GTSRB dataset is an image  collection
  consisting of  43 traffic signs~\cite{Stallkamp2012}. Images vary in size and are
RGB-encoded. To simplify, we resize images to $32$x$32$
pixels, recenter them by subtracting the mean component, and rescale
them by factoring their standard deviations out. We keep $35,000$ images for
our training set and $4,000$ for our validation set (out of the $39,209$
available), and $10,000$ for our test set (out of $12,630$). We train the
oracle on our machine, using the DNN  B from
Table~\ref{table:dnn-architectures} (cf. appendix), for $50$ epochs with a learning rate
of $10^{-2}$ and a momentum of $0.9$ (both decayed by $0.5$ every $10$
epochs).

\boldpara{Substitute DNN Training} The adversary uses two 
initial substitute training sets extracted from the GTSRB test set. The
first includes the first $1,000$ samples and the second the
first $500$. The number of initial samples is higher than for 
MNIST substitutes as inputs have a higher dimensionality. We train 
three substitute architectures C, D, and E (cf.
Table~\ref{table:dnn-architectures}) using the oracle for $6$ substitute
training epochs with a Jacobian-based dataset augmentation parameter of
$\lambda=0.1$. Substitute C and E where trained with the $1,000$ sample
initial substitute training set and achieve a $71.42\%$ accuracy. Substitute D
was trained with the initial set of $500$ samples. Its accuracy of $60.12\%$ is
lower than C and E.

\begin{figure}[t]	
	\includegraphics[width=\columnwidth]{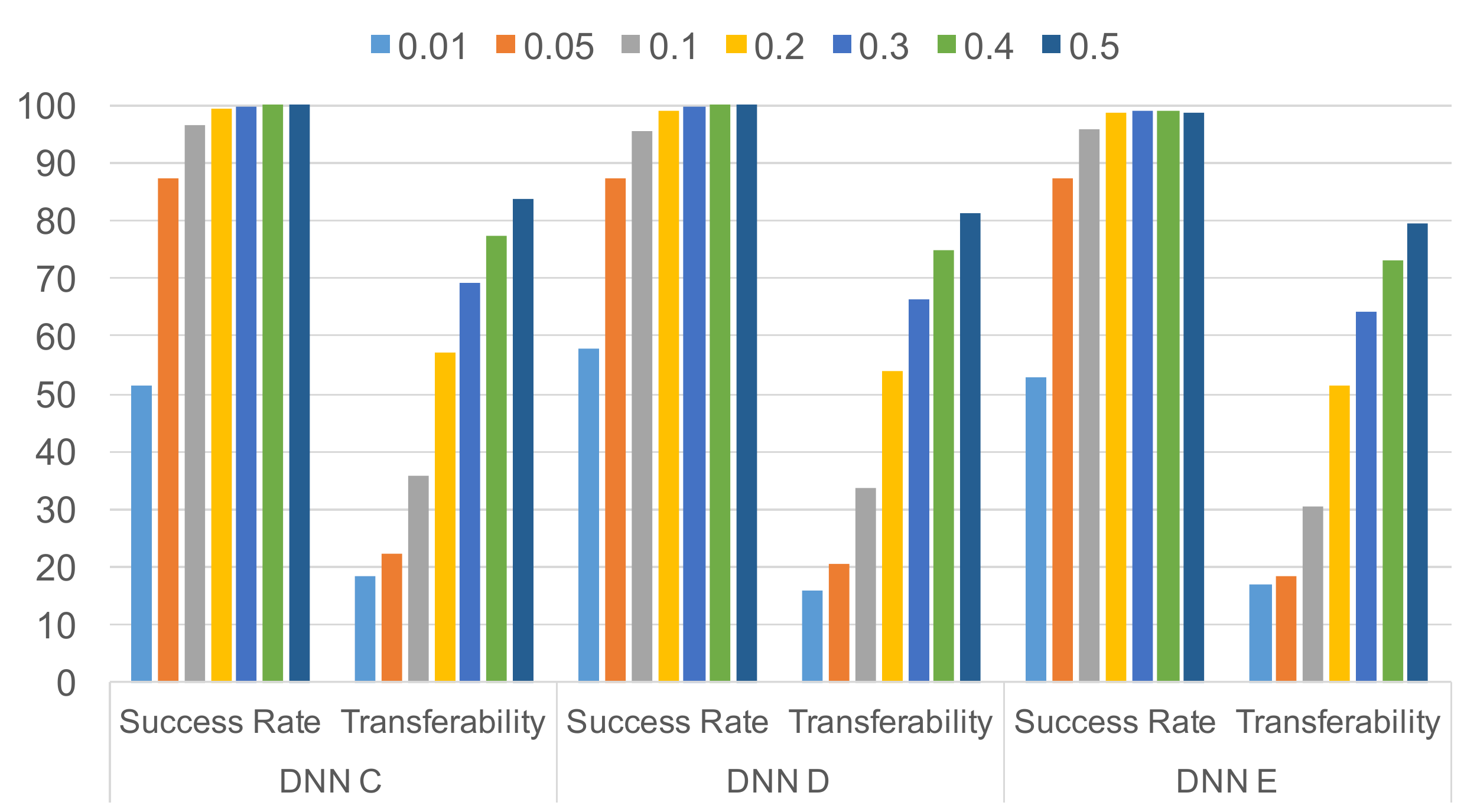}
	\caption{\textbf{Success Rate and Transferability of Adversarial Samples crafted on the GTRSRB dataset: }each bar corresponds to a different input variation.}
	\label{fig:gtsdb-transferability}
\end{figure}

\boldpara{Adversarial Crafting} We use Goodfellow's algorithm
with $\varepsilon$ between $0.01$ and $0.5$ to craft
adversarial samples from the test set. Results 
are shown in Figure~\ref{fig:gtsdb-transferability}. Adversarial samples
crafted with variations $\varepsilon<0.3$ are more transferable than
those crafted with the same $\varepsilon$ for MNIST models. This is likely due to the higher
input dimensionality---$3,072$ components instead of $784$---which means almost $4$ times
 more perturbation is applied with the same $\varepsilon$.
Nevertheless, with success rates higher than $98.98\%$ and transferability
rates ranging from $64.24\%$ to $69.03\%$ for $\varepsilon=0.3$, which is hard
to distinguish for humans, \emph{the attack is successful}.  The
transferability of adversarial samples crafted using substitute DNN D is
comparable or higher than corresponding samples for DNNs C and E,
despite being less
accurate (trained with less samples). This emphasizes that there is no strong correlation between substitute accuracy and transferability.

\section{Attack Algorithm Calibration}
\label{sec:evaluation}

Having shown in Section~\ref{sec:validation} that an adversary can force an
MNIST oracle from MetaMind, and a GTSRB oracle trained locally, to
misclassify inputs,
we now perform a parameter space exploration of both attack steps--the
substitute DNN training and the adversarial sample crafting. We explore the following questions: ``(1) How can substitute
training be fine-tuned to improve adversarial sample transferability?'' and (2)
``For each adversarial sample crafting strategies, which parameters optimize
transferability?''. We found that:
\vspace*{-0.05in}
\begin{itemize}[leftmargin=0.25in]
\setlength\itemsep{0.01em}
\item In Section~\ref{ssec:substitute-dnn-training}, we show that the choice of substitute DNN architecture (number of layers, size,
activation function, type) has a limited impact on adversarial sample
transferability. Increasing the number of epochs, after the substitute DNN has
reached an asymptotic accuracy, does not improve adversarial sample
transferability. 
\item At comparable input perturbation magnitude, the Goodfellow and Papernot
algorithms have similar transferability rates (see  Section~\ref{ssec:adv-sample-crafting-comparison}). 
\end{itemize}
\vspace*{-0.05in}
In this section, we use an oracle trained locally to limit querying of the
MetaMind API. We train architecture A (cf. Table~\ref{table:dnn-architectures})
for $50$ epochs with a learning parameter $10^{-2}$ and a momentum $0.9$ (both
decayed by $0.5$ every $10$ epochs). 

\subsection{Calibrating Substitute DNN Training}
\label{ssec:substitute-dnn-training}

We first seek to quantify the impact of substitute training algorithm parameters on adversarial sample transferability and introduce a refinement to reduce oracle querying.


\boldpara{Choosing an Architecture}  We train substitute DNNs A and F to M (cf.
Table~\ref{table:dnn-architectures}) using $150$ samples from the MNIST test set as the substitute training set. During each
of the $6$ substitute training epochs, the DNN is trained for $5$ epochs
from scratch. Between epochs, synthetic data is added to the training set using Jacobian-based
dataset augmentations with step $\lambda = 0.1$. The substitute
architectures differ from the oracle's by the type, number, and size
of layers. In Table~\ref{table:goodfellow-transferability-architectures},
we report the accuracy of each architecture after $2$ and $6$ substitute training epochs, as well as the adversarial
sample transferability after 6 epochs. Adversarial samples are crafted using the Goodfellow algorithm with an input
variation of $\varepsilon=0.4$ (which we justify later). The last column of
Table~\ref{table:goodfellow-transferability-architectures} 
shows that the choice of architecture has a limited
impact on adversarial sample transferability, and therefore on the attack
success. The most important transferability drop follows from removing all
convolutional layers. Changing the hidden layer activation function from
rectified linear to a sigmoid does not impact transferability significantly.


\boldpara{Choosing the number of substitute epochs} Another
tunable parameter is the number of
epochs for which substitute DNNs are trained. Intuitively, one would hypothesize that
the longer we train the substitute, the more samples labeled using the
oracle are included in the substitute training set, thus the higher the
transferability of adversarial samples will be. This intuition is confirmed
only partially by our experiments on substitute DNN A. We find that for
for input variations $\varepsilon\leq 0.3$, the transferability is slightly
improved by a rate between $+3\%$ to $+9\%$, but for variations
$\varepsilon\geq 0.4$, the transferability is slightly degraded by less than
$1\%$.

{

\begin{table}[t]
	\centering
	\begin{tabular}{|c|c |c| c|}
		\hline
		DNN & Accuracy & Accuracy & Transferability \\ 
		ID & ($\rho =2$) & ($\rho =6$) & ($\rho =6$) \\ \hline \hline
		A & $30.50\%$ &  $82.81\%$& $75.74\%$ \\ \hline 
		F & $68.67\%$ &  $79.19\%$ &  $64.28\%$ \\ \hline 
		G & $72.88\%$ &  $78.31\%$ &  $61.17\%$ \\ \hline 
		H & $56.70\%$ &  $74.67\%$ &  $63.44\%$ \\ \hline 
		I & $57.68\%$ &  $71.25\%$ &  $43.48\%$ \\ \hline 
		J & $64.39\%$ &  $68.99\%$ &  $47.03\%$ \\ \hline 
		K & $58.53\%$ &  $70.75\%$ &  $54.45\%$ \\ \hline 
		L & $67.73\%$ &  $75.43\%$ &  $65.95\%$ \\ \hline 
		M & $62.64\%$ &  76.04 & $62.00\%$\\ \hline 
	\end{tabular}
	\caption{\textbf{Substitute Accuracy} at $\rho=2$ and $\rho =6$ substitute epochs and \textbf{Transferability of Adversarial Samples:} for $\varepsilon=0.4$ after $\rho =6$ substitute epochs.}
	\label{table:goodfellow-transferability-architectures}
\end{table}}

\boldpara{Setting the step size} We trained 
substitute A using different Jacobian-based
dataset augmentation step sizes $\lambda$. Increasing or decreasing the step size (from $\lambda=0.1$ used in the 
rest of this paper) does not modify the substitute accuracy by more than $3\%$. Larger step sizes decrease convergence stability while smaller values yield slower convergence.
However, increasing step size $\lambda$ negatively
impacts adversarial sample transferability : for instance with a step size 
of $0.3$ compared to $0.1$, the transferability rate for $\varepsilon=0.25$
is $10.82\%$ instead of $22.35\%$ and for $\varepsilon=0.5$, $82.07\%$
instead of $85.22\%$.

However, having the step size periodically alternating between positive and negative
values improves the quality of the oracle approximation made by the substitute. This could be explained by the fact that after a few substitute epochs, synthetic inputs are outside of the input domain and are thus clipped to produce an acceptable input.
We introduce an iteration period $\tau$
after which the step size is multiplied by $-1$. Thus, the step size
$\lambda$ is now replaced by:
\begin{equation}
\label{fig:periodical-step-size}
\lambda_\rho = \lambda \cdot (-1)^{\left\lfloor \frac{\rho}{\tau} \right\rfloor}
\end{equation}
where $\tau$ is set to be the number of epochs after which the Jacobian-based
dataset augmentation does not lead any substantial improvement in the
substitute. A grid search can also be performed to find an optimal value for
the period $\tau$. We also experimented with a decreasing grid step amplitude
$\lambda$, but did not find that it yielded substantial improvements. 

\boldpara{Reducing Oracle Querying} We apply \emph{reservoir sampling}~\cite{vitter1985random} 
 to reduce the number of queries made to the oracle. This is useful
when learning substitutes in realistic environments, or when interacting with paid APIs, where the number of label queries an adversary can make without
exceeding a quota or being detected by a defender is
limited. Reservoir
sampling is a technique that randomly select $\kappa$ samples from a list of samples. 
The
total number of samples in the list can be both very large and unknown. 
We use it to select $\kappa$ new inputs
 before a Jacobian-based dataset augmentation. This prevents
the exponential growth of queries made to the oracle at each augmentation. At iterations $\rho>\sigma$ (the first $\sigma$ iterations are
performed normally), when considering the previous set $S_{\rho-1}$ of
substitute training inputs, we select $\kappa$ inputs from $S_{\rho-1}$ to be
augmented in $S_\rho$. Using reservoir
sampling ensures that each input in $S_{\rho-1}$ has an equal probability
$\frac{1}{\left| S_{\rho-1} \right|}$ to be augmented in $S_\rho$. The number
of queries made to the oracle is reduced from $n\cdot2^\rho$ for the vanilla
Jacobian-based augmentation to $n\cdot2^\sigma+\kappa\cdot(\rho-\sigma)$  with reservoir sampling. In Section~\ref{sec:generalization}, we show that using reservoir sampling to reduce
the number of synthetic training inputs does not significantly degrade the substitute accuracy.

\subsection{Adversarial Sample Crafting}
\label{ssec:adv-sample-crafting-comparison}

We compare the transferability of
adversarial samples produced by each algorithm introduced
previously~\cite{goodfellow2014explaining,
papernot2015limitations}, to elect the strongest technique under our threat model.

\begin{figure}[t]
	\includegraphics[width=\columnwidth]{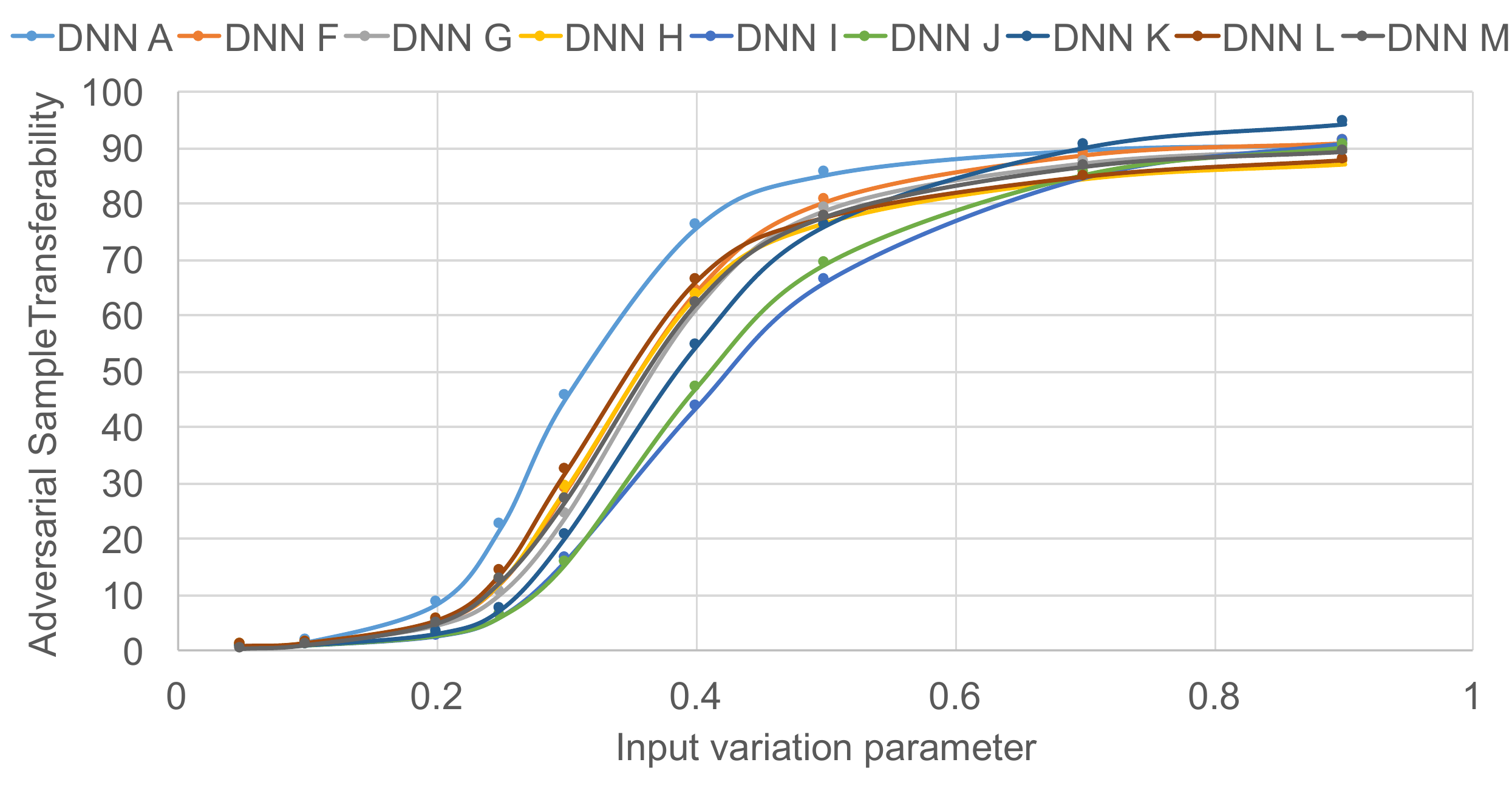}
	\caption{\textbf{Impact of input variation $\varepsilon$ in the Goodfellow crafting algorithm on the transferability of adversarial samples:} for architectures from Table~\ref{table:goodfellow-transferability-architectures}.}
	\label{fig:goodfellow-factor-transferability}
	\vspace*{-0.15in}
\end{figure}

\boldpara{Goodfellow's algorithm} Recall from Equation~\ref{eq:goodfellow-perturbation} the perturbation computed in the Goodfellow attack.
Its only parameter is the variation $\varepsilon$ added in
the direction of the gradient sign. We use the same architecture set as
before to quantify the impact of $\varepsilon$ on 
adversarial sample transferability. 
In Figure~\ref{fig:goodfellow-factor-transferability}, architecture A outperforms
all others: it is a copy of the oracle's and acts as a baseline. Other
architectures have asymptotic transferability rates ranging between $72.24\%$
and $80.21\%$, confirming that \emph{the substitute architecture choice has
a limited impact on transferability}. Increasing the value of $\varepsilon$ above
$0.4$ yields little improvement in transferability and should be avoided
to guarantee indistinguishability of adversarial samples to humans.


\boldparab{Papernot's algorithm} This algorithm is fine-tuned by
two parameters: the \emph{maximum distortion} $\Upsilon$ and the \emph{input
variation} $\varepsilon$. The maximum distortion\footnote{In~\cite{papernot2015limitations}, the algorithm stopped perturbing when the input reached the target class. Here, we force
the algorithm to continue perturbing until it changed
$\Upsilon$ input components.} defines the number of input components that are altered in
perturbation $\delta_{\vec{x}}$. The input variation,
similarly to the Goodfellow algorithm, controls the amount of change induced to
altered input components.

\begin{figure}[t]
	\centering
	\includegraphics[width=0.9\columnwidth]{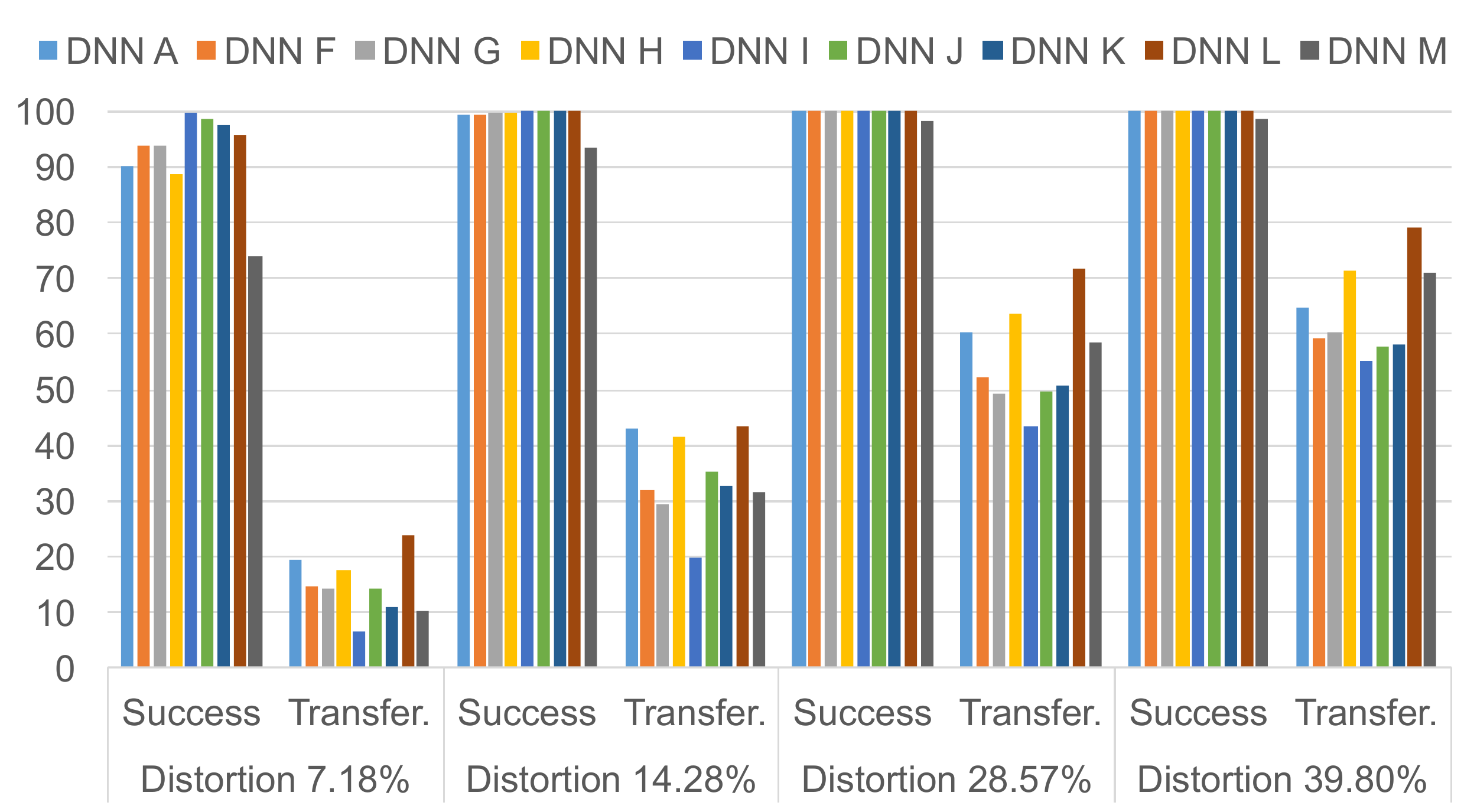}
	\caption{\textbf{Impact of the maximum distortion $\Upsilon$ in the Papernot  algorithm on success rate and transferability of adversarial samples:} increasing $\Upsilon$ yields higher transferability rates across DNNs.}
	\label{fig:papernot-max-distortion}
\end{figure}

We first evaluate the impact of the maximum distortion $\Upsilon$ on
adversarial sample transferability. For now, components selected to be
perturbed are increased by $\varepsilon=1$. Intuitively, 
increasing the maximum distortion makes adversarial samples more
transferable. Higher distortions
increase the misclassification confidence of the substitute DNN, and
also increases the likelihood of the oracle misclassifying the same sample. These results are reported in
Figure~\ref{fig:papernot-max-distortion}. Increasing distortion $\Upsilon$ from
$7.14\%$ to $28.57\%$ improves transferability: at a $7.14\%$ distortion, the
average transferability across all architectures is $14.70\%$
whereas at a $28.57\%$ distortion, the average transferability is at $55.53\%$.

\begin{figure}[t!]
	\centering
	\includegraphics[width=0.9\columnwidth]{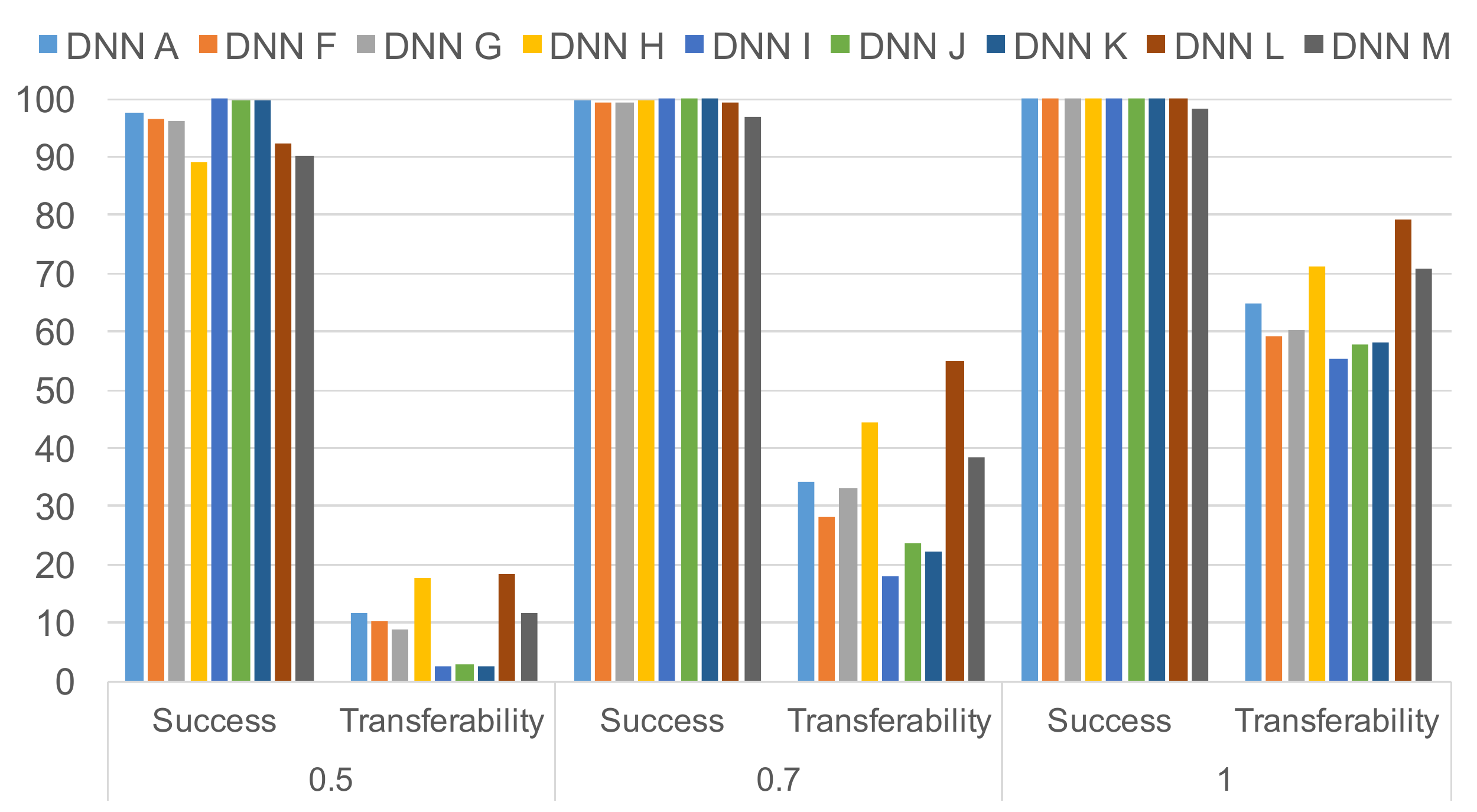}
	\caption{\textbf{Impact of the input variation $\varepsilon$ in the Papernot algorithm on the success rate and adversarial sample transferability} computed for $\varepsilon\in \{0.5, 0.7, 1\}$ on DNNs from Table~\ref{table:goodfellow-transferability-architectures} with distortion  $\Upsilon=39.80\%$.}
	\label{fig:papernot-input-variation}
	\vspace*{-0.15in}
\end{figure}

We now quantify the impact of the variation
$\varepsilon$ introduced to each input component selected in
$\delta_{\vec{x}}$. We find that reducing the input variation 
from $1$ to $0.7$
significantly degrades adversarial sample transferability,
approximatively by a factor of 2 (cf. Figure~\ref{fig:papernot-input-variation}). This is explained by the fixed
distortion parameter $\Upsilon$, which prevents the crafting algorithm 
from increasing the number of components altered to compensate for the reduced effectiveness yielded by the smaller $\varepsilon$. 

\boldpara{Comparing Crafting Algorithms} To compare the two crafting strategies and their differing perturbation styles fairly,
we compare their success rate given a fixed L1 norm of the introduced perturbation $\delta_{\vec{x} }$, which can be defined as:
\begin{equation}
\| \delta_{\vec{x}} \|_1 = \varepsilon \cdot \|  \delta_{\vec{x}} \|_0
\end{equation}
where $\|  \delta_{\vec{x}} \|_0$ is the number of input components 
selected in the perturbation $\delta_{\vec{x}}$,
 and $\varepsilon$ the input
variation introduced to each component perturbed. 
For the Goodfellow algorithm, we always
have $\| \delta_{\vec{x}} \|_0=1$, whereas for the Papernot algorithm, values vary
 for both $\varepsilon$ and $ \|\delta_{\vec{x}}\|_0$. For instance, 
  $\| \delta_{\vec{x}} \|_1 =0.4$ corresponds to a
Goodfellow algorithm with $\varepsilon=0.4$ and a Papernot algorithm with
$\varepsilon=1$ and  $\Upsilon=40\%$. Corresponding transferability
rates can be found in
Table~\ref{table:goodfellow-transferability-architectures} and
Figure~\ref{fig:papernot-max-distortion} for our running set of architectures.
Performances are comparable with some DNNs performing better with one algorithm and others with the other. 
Thus, the choice of algorithm depends on acceptable perturbations: 
e.g., all features perturbed a little vs. few features perturbed a lot. 
Indeed, the Goodfellow algorithm gives more control on $\varepsilon$ while the 
Papernot algorithm gives more control on $\Upsilon$.


\section{Generalization of the Attack}
\label{sec:generalization}

\vspace*{-0.05in}

So far, all substitutes and oracles considered were learned with DNNs.
However, no part of the attack limits its applicability to other ML techniques.
For instance, we show that the attack generalizes to non-differentiable \emph{target oracles} like decision trees.
As pointed out by Equation~\ref{eq:jacobian-augmentation}, the only limitation
is placed on the \emph{substitute}: it must model a differentiable
function---to allow for synthetic data to be generated with its Jacobian matrix.
We show below that:
\vspace*{-0.05in}
\begin{itemize}
	\setlength\itemsep{-0.1cm}
	\item Substitutes can also be learned with logistic regression.
	\item The attack generalizes to additional ML models by:
	(1) learning substitutes of 4 classifier types
	(logistic regression, SVM, decision tree, nearest neighbors) in addition to DNNs,
	and (2) targeting remote models hosted by Amazon Web Services and Google Cloud
	Prediction with success rates of $96.19\%$ and $88.94\%$ after $800$ queries
	to train the substitute.   
\end{itemize}

\subsection{Generalizing Substitute Learning} 
\label{sec:dnn-approximators}

We here show that our approach generalizes to ML models that are not DNNs.
Indeed, we learn substitutes for 4 representative types of ML classifiers in
addition to DNNs: logistic regression (LR), support vector machines (SVM),
decision trees (DT), and nearest neighbor (kNN). All of these classifiers are
trained on MNIST, with no feature engineering (i.e. directly on raw pixel values) as done in Section~\ref{sec:validation}.

Whereas we previously trained all of our substitutes using DNNs only, we now use
both DNNs and LR as substitute models. The Jacobian-based dataset augmentation
described in the context of DNNs is easily adapted to logistic regression: the
later is analog to the softmax layer frequently used by the former when
outputting probability vectors. We use $100$ samples from the MNIST test set as
the initial substitute training set and use the two refinements introduced in
Section~\ref{sec:evaluation}: a \emph{periodic step size} and \emph{reservoir
	sampling}.

\begin{figure}[p] 
	\begin{subfigure}[b]{\columnwidth}
		\centering
		\includegraphics[width=0.85\columnwidth]{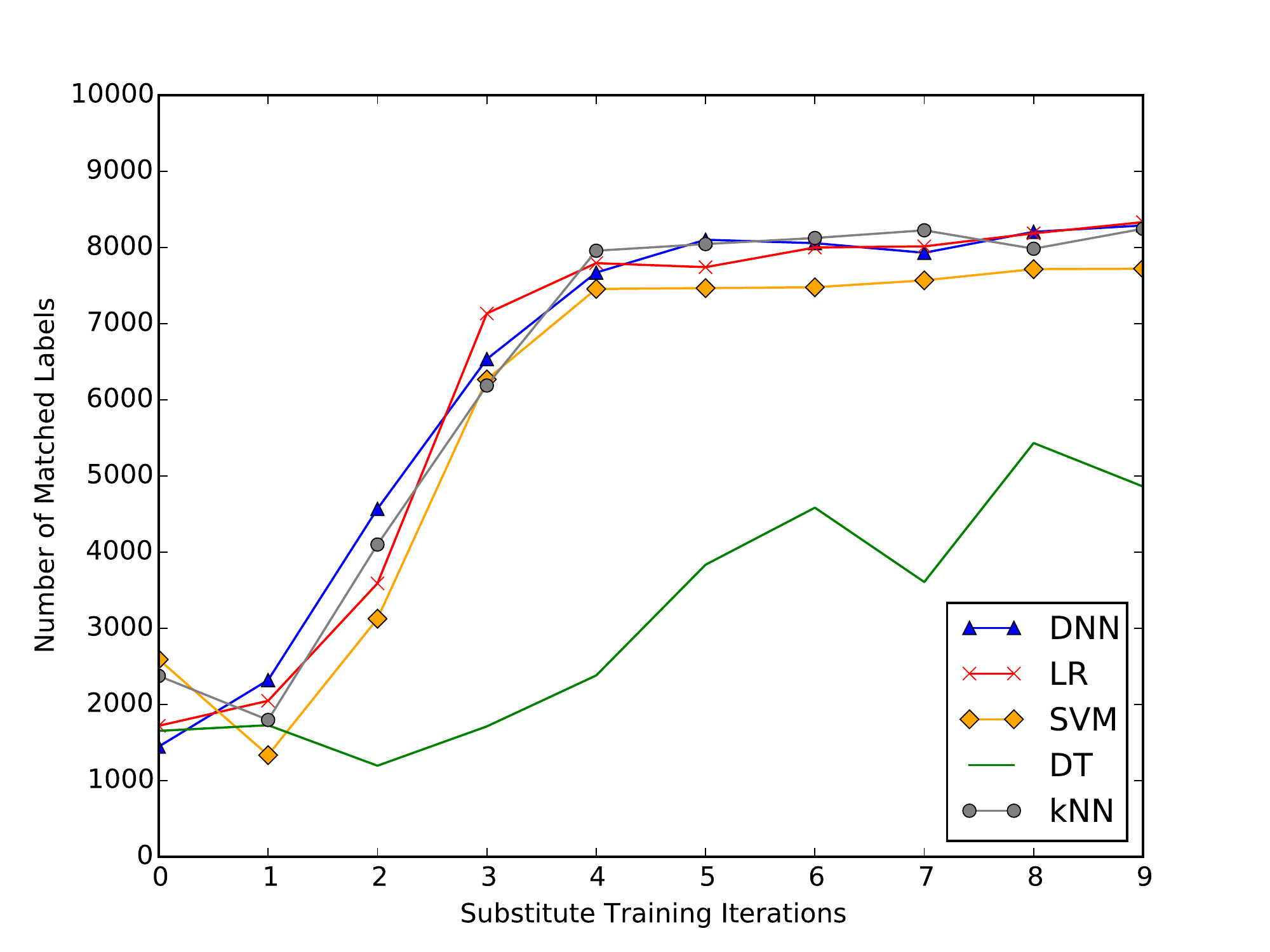} 
		\caption{DNN substitutes} 
		\label{fig:learning-approximators-dnn} 
	\end{subfigure}
	\begin{subfigure}[b]{\columnwidth}
		\centering
		\includegraphics[width=0.85\columnwidth]{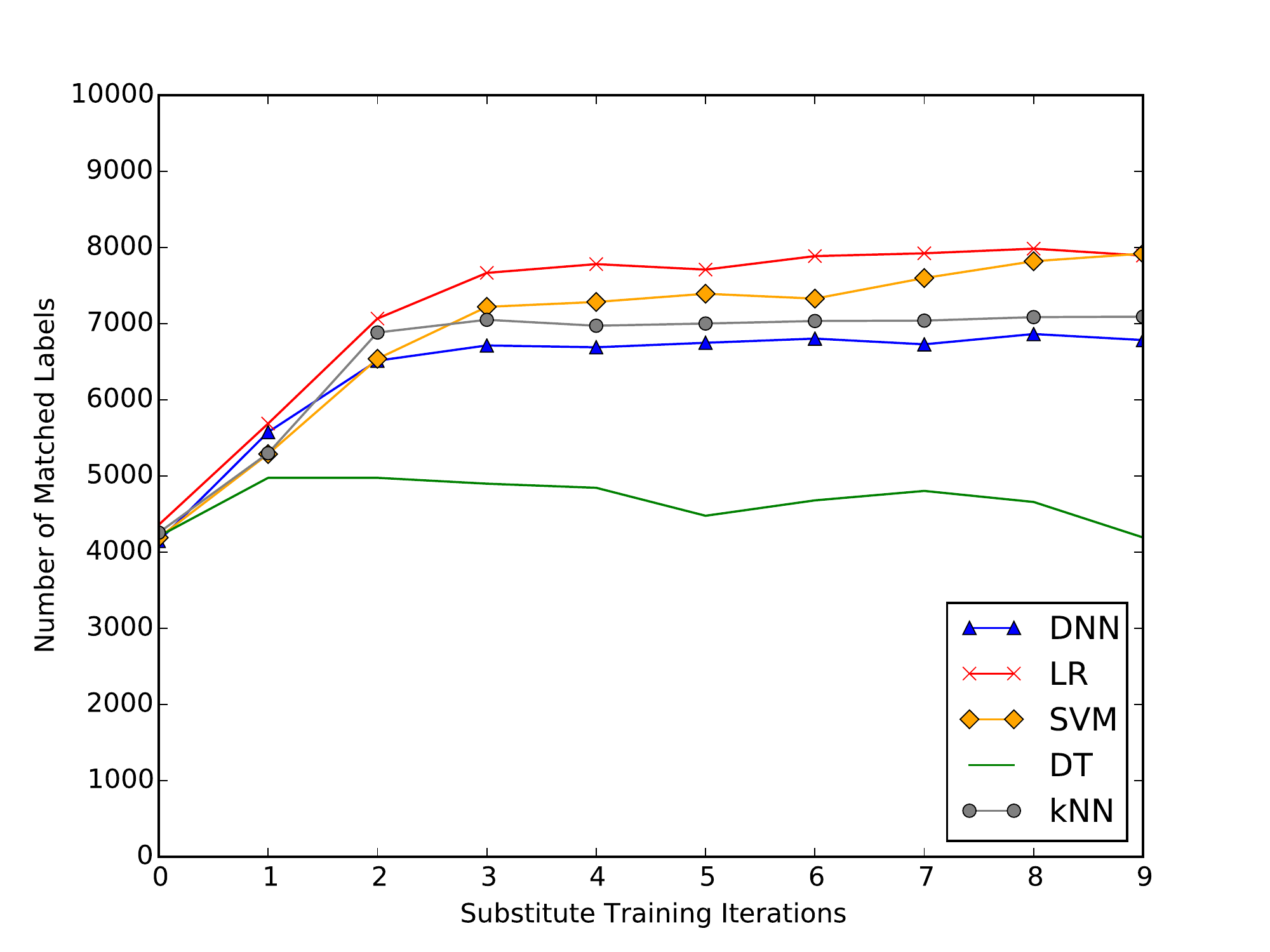} 
		\caption{LR substitutes} 
		\label{fig:learning-approximators-lr} 
	\end{subfigure}
	\caption{\textbf{Label predictions matched} between the  substitutes (DNN and LR) and their target  oracles on test data.}
	\label{fig:learning-approximators} 
\end{figure}

Figure~\ref{fig:learning-approximators}(a) and~\ref{fig:learning-approximators}(b)  plot for each iteration $\rho$ the
share of samples on which the substitute DNNs and LRs agree with predictions made by
the oracle they are approximating. This proportion is estimated by
comparing labels assigned to the test set by the substitutes and
oracles before each iteration $\rho$ of the Jacobian-based dataset
augmentation. 
All substitutes are able to 
approximate the corresponding oracle at rates higher between $77\%$ and $83\%$  after $\rho=10$ iterations (to the exception of the decision tree oracle, which could be due to its non-continuity). 
LR substitute accuracies are
generally lower than those of DNN substitutes, except when targeting the LR and SVM oracles where LR substitutes outperform DNN ones. However, LR substitutes are computationally more efficient and reach their asymptotic match rate faster,
after $\rho=3$ iterations, corresponding to $800$ oracle
queries.

Table~\ref{tbl:refinements} quantifies the impact of refinements introduced in Section~\ref{sec:evaluation} on results reported in Figure~\ref{fig:learning-approximators}(a) and~\ref{fig:learning-approximators}(b). 
The \emph{periodic step size} (PSS) increases the oracle approximation accuracy of
substitutes. After
$\rho=9$ epochs, a substitute DNN trained with PSS
matches $89.28\%$ of the DNN oracle labels, whereas the vanilla substitute 
DNN matches only $78.01\%$.
Similarly, the LR substitute with PSS matches $84.01\%$ of the LR oracle labels while the vanilla substitute matched $72.00\%$. Using \emph{reservoir sampling} (RS) reduces oracle querying. For
instance, $10$ iterations with RS ($\sigma=3$ and $\kappa
=400$) make $100\cdot 2^3 + 400(10-3)=3,600$ queries to the oracle instead of
$102,400$ without RS. This decreases the substitute accuracy,
but when combined with PSS it remains superior to the vanilla substitutes. For instance, the vanilla substitute matched $7,801$ of the DNN oracle labels,
the PSS one $8,928$, and the PSS with RS one $8,290$. Simarly,
 the vanilla LR substitute matched $71.56\%$ of the SVM oracle labels, the PSS one $82.19\%$, and the PSS with RS $79.20\%$.

\begin {table}[p]
\centering
\begin{footnotesize}
\begin{tabular}{|l||c|c|c|c|c|}
	\hline
	Substitute  & DNN & LR & SVM & DT & kNN     \\  \hline \hline
	DNN & 78.01  & 82.17  & 79.68  & 62.75  & 81.83  \\ \hline
	DNN+PSS & \textbf{89.28}  & \textbf{89.16}  & \textbf{83.79}   & \textbf{61.10}  & \textbf{85.67}  \\ \hline
	DNN+PSS+RS & 82.90  & 83.33   & 77.22  & 48.62  & 82.46   \\ \hline\hline
	LR & 64.93  & 72.00  & 71.56  & 38.44  & 70.74   \\ \hline
	LR+PSS & \textbf{69.20} & \textbf{84.01}  & \textbf{82.19}  & \textbf{34.14}  & \textbf{71.02} \\ \hline
	LR+PSS+RS & 67.85  & 78.94  & 79.20  & 41.93  & 70.92  \\ \hline
	\hline
\end{tabular}
\end{footnotesize}
\caption{\textbf{Impact of our refinements}, Periodic Step Size (PSS) and Reservoir Sampling (RS), on the percentage of label predictions matched between the substitutes and their target classifiers on test data  after $\rho=9$ substitute iterations. }
\label{tbl:refinements}
\end{table}

\begin {table}[p]
\centering
\begin{footnotesize}
\begin{tabular}{|c|c||c|c|c|c|}
	\hline
	\multicolumn{2}{|c||}{}   & \multicolumn{2}{c|}{Amazon}  & \multicolumn{2}{c|}{Google}    \\  \hline
	Epochs & Queries & DNN & LR  & DNN & LR    \\  \hline \hline
	$\rho=3$ & 800 & 87.44 &  96.19 & 84.50&  88.94 \\ \hline
	$\rho=6$ & 6,400  & \textbf{96.78} & \textbf{96.43}  & \textbf{97.17} & 92.05  \\ \hline
	$\rho=6^*$ & 2,000 & 95.68& 95.83& 91.57& \textbf{97.72}   \\ \hline
\end{tabular}
\end{footnotesize}
\caption{\textbf{Misclassification rates (\%) of the Amazon and Google oracles} on adversarial samples produced with DNN and LR substitutes after $\rho=3,6$ epochs. The 2nd column is the number of queries during substitute training. Last row uses  a periodic step size and reservoir sampling.}
\label{tbl:aws-misclassification}. 
\end{table}

\subsection{Attacks against Amazon \& Google oracles}


\boldpara{Amazon oracle} To train a classifier on \emph{Amazon Machine
	Learning},\footnote{\url{https://aws.amazon.com/machine-learning}}, we  uploaded a CSV version of the MNIST
dataset to a S3 bucket. 
We then loaded the
data, selected the multi-class model type, and keept default configuration settings. The process took a few minutes and produced a classifier achieving a
$92.17\%$ test set accuracy. We cannot improve the accuracy due to the automated nature of training. We then activate real-time predictions to query the model for labels from our machine with the provided API. 
Although probabilities are returned, we discard them and retain
\emph{only the most likely label}---as stated in our threat model
(Section~\ref{ssec:threat-model}).

\boldpara{Google oracle} The procedure
to train a classifier on Google's Cloud Prediction API\footnote{\scriptsize{\url{https://cloud.google.com/prediction/}}} is similar to Amazon's. We 
upload the CSV file with the MNIST
training data to Google Cloud Storage. 
We then train a model using the Prediction API.
The only property we can
specify is the expected multi-class nature of our model.
We then evaluate the resulting
model on the MNIST test set.
The API  reports
an accuracy of $92\%$ on this test set for the model trained.

\boldpara{Substitute Training} By augmenting an initial training set of $100$ test set samples, we
train a DNN and LR substitute for each of the two oracles. We measure success  as the rate of adversarial
samples misclassified by the corresponding oracle, among the $10,000$ produced from the test set using the fast gradient sign method with parameter $\varepsilon=0.3$. These rates, computed after $\rho\in\{3,6\}$ dataset augmentation iterations, are reported in Table~\ref{tbl:aws-misclassification}. Results reported in the last row use both a periodic step size and reservoir sampling (hence the reduced number of queries made to train the substitute).

\boldpara{Experimental Results}  With a $96.19\%$ misclassification
rate for a perturbation $\varepsilon=0.3$ crafted using a LR substitute
trained with $800$ oracle queries, the model hosted by Amazon is easily
misled. The model trained by Google is somewhat more
robust to adversarial samples, but
is still vulnerable to a large proportion of samples: $88.94\%$ of adversarial
samples produced in the same conditions are misclassified. A careful read of the  documentation indicated that the model trained by Amazon is a multinomial logistic regression.\footnote{{\scriptsize\url{docs.aws.amazon.com/machine-learning}}}
As pointed out in~\cite{goodfellow2014explaining}, shallow models like logistic regression
are unable to cope with adversarial samples and learn robust classifiers. This explains why the attack is very successful and the LR
substitute performs better than the DNN substitute. We were however not able to find the ML technique Google uses.

The last row of Table~\ref{tbl:aws-misclassification} shows how combining periodic step sizes with
reservoir sampling allow us to reduce querying of both oracles during
substitute training, while crafting adversarial samples with 
higher transferability to the target classifier. 
Indeed, querying is reduced by a factor larger than $3$ from $6,400$ to $2,000$ queries,
while misclassification decreases only from $96.78\%$ to $95.68\%$ for the Amazon DNN substitute.
It is still larger than 
the rate of $87.44\%$ achieved after $800$ queries by the substitute learned without the refinements. 
Similarly, the 
misclassification rate of the Google LR substitute is $97.72\%$---compared to $92.05\%$ with the original method after $\rho=6$ epochs, confirming the result.


\section{Defense Strategies}
\label{sec:defenses}

The two types of defense strategies are: (1) \emph{reactive} where one seeks to
detect adversarial examples, and (2) \emph{proactive} where one makes the model
itself more robust. Our attack is not more easily detectable
than a classic adversarial example attack. Indeed, oracle queries may be
distributed among a set of colluding users, and as such remain hard to detect. The defender may increase the attacker's cost by training models with
higher input dimensionality or modeling complexity, as our
experimental results indicate that these two factors increase the number of queries required
to train  substitutes. In the following, we thus only analyze our attack in the face of defenses that seek to make the
(oracle) model robust.

Many potential defense mechanisms fall into a category we call {\em gradient
	masking}. These techniques construct a model that does not have useful
gradients, e.g., by using a nearest neighbor classifier instead of a DNN. Such
methods make it difficult to construct an adversarial example directly, due to
the absence of a gradient, but are often still vulnerable to the adversarial
examples that affect a smooth version of the same model. Previously, it has been
shown that nearest neighbor was vulnerable to attacks based on transferring
adversarial examples from smoothed nearest
neighbors\cite{goodfellow2014explaining}.

We show a more general flaw in the category of gradient masking.
Even if the defender attempts to prevent attacks by not publishing
the directions in which the model is sensitive, these directions 
can be discovered by other means, in which case the
same attack can still succeed.
We show that the black-box attack based on transfer from a substitute model
overcomes gradient masking defenses. No fully effective defense mechanism is known, but we study the two with the
greatest empirical success so far:
adversarial training~\cite{goodfellow2014explaining,szegedy2013intriguing}, and
defensive distillation for DNNs~\cite{papernot2015distillation}.

\begin {table}[t]
\centering
	\begin{tabular}{|c|c|c|c|c|}
		\hline
		Training $\varepsilon$   & Attack $\varepsilon$ & O$\rightarrow$O  & S $\rightarrow$ S & S $\rightarrow$ O   \\  \hline
		$0.15$  & $0.3$ & $10.12\%$  & $94.91\%$ & $38.54\%$   \\  \hline
		$0.15$ & $0.4$  & $43.29\%$ & $99.75\%$ &   $71.25\%$ \\  \hline
		$0.3$ & $0.3$ & $0.91\%$ & $93.55\%$ &  $1.31\%$  \\  \hline
		$0.3$ & $0.4$ & $29.56\%$ & $99.48\%$ &  $10.30\%$  \\  \hline
	\end{tabular}
\caption{\textbf{Evaluation of adversarial training}:
	the columns indicate the input variation parameter used to
	inject adversarial examples during training and to compute the attacks,
	the attack success rate when examples crafted on the (O)racle are deployed against the (O)racle,
	the attack success rate when examples crafted on the (S)ubstitute are deployed against the (S)ubstitute,
	and
	the attack success rate when examples crafted on the (S)ubstitute are deployed against the (O)racle.}
\label{tbl:adv-training}. 
\vspace*{-0.3in}
\end{table}

\boldpara{Adversarial training}
It was shown that injecting adversarial examples throughout training increases
the robustness of significantly descriptive models, such as DNNs~\cite{goodfellow2014explaining,szegedy2013intriguing,warde2016adversarial}.
We implemented an approximation of this defense using the Google Prediction API.
Since the API does not support the generation of adversarial examples
at every step of training, as a correct implementation of adversarial training would
do, we instead inject a large amount of adversarial examples infrequently.
After training in this way, the model has a misclassification rate of $8.75\%$ on
the unperturbed test set,
but the adversarial misclassification rate rises to $100\%$ when $\rho=6$.
To evaluate this defense strategy using a correct implementation, we resort
to training the oracle locally, using our own codebase that includes support for
generating adversarial examples at each step.
After each training batch, we compute and train on adversarial examples
generated with the fast gradient sign method before starting training on the next batch of the
original training data.
Results are given in Table~\ref{tbl:adv-training}.
We observe that for $\varepsilon=0.15$, the defense can be evaded using the
black-box attack with adversarial examples crafted on the substitute and
misclassified by the oracle at rates up to $71.25\%$.
However, for $\varepsilon=0.3$, the black-box attack is not effective anymore.
Therefore, making a machine learning model robust to small and infinitesimal
perturbations of its inputs is an example of \emph{gradient masking} and can
be evaded using our substitute-based black-box approach.
However, making the model robust to larger and finite perturbations prevents
the black-box attack.
To confirm this hypothesis, we now show that defensive distillation, which
makes the model robust to infinitesimal perturbations, can be evaded by the
black-box approach.

\begin{figure}[t]
	\includegraphics[width=\columnwidth]{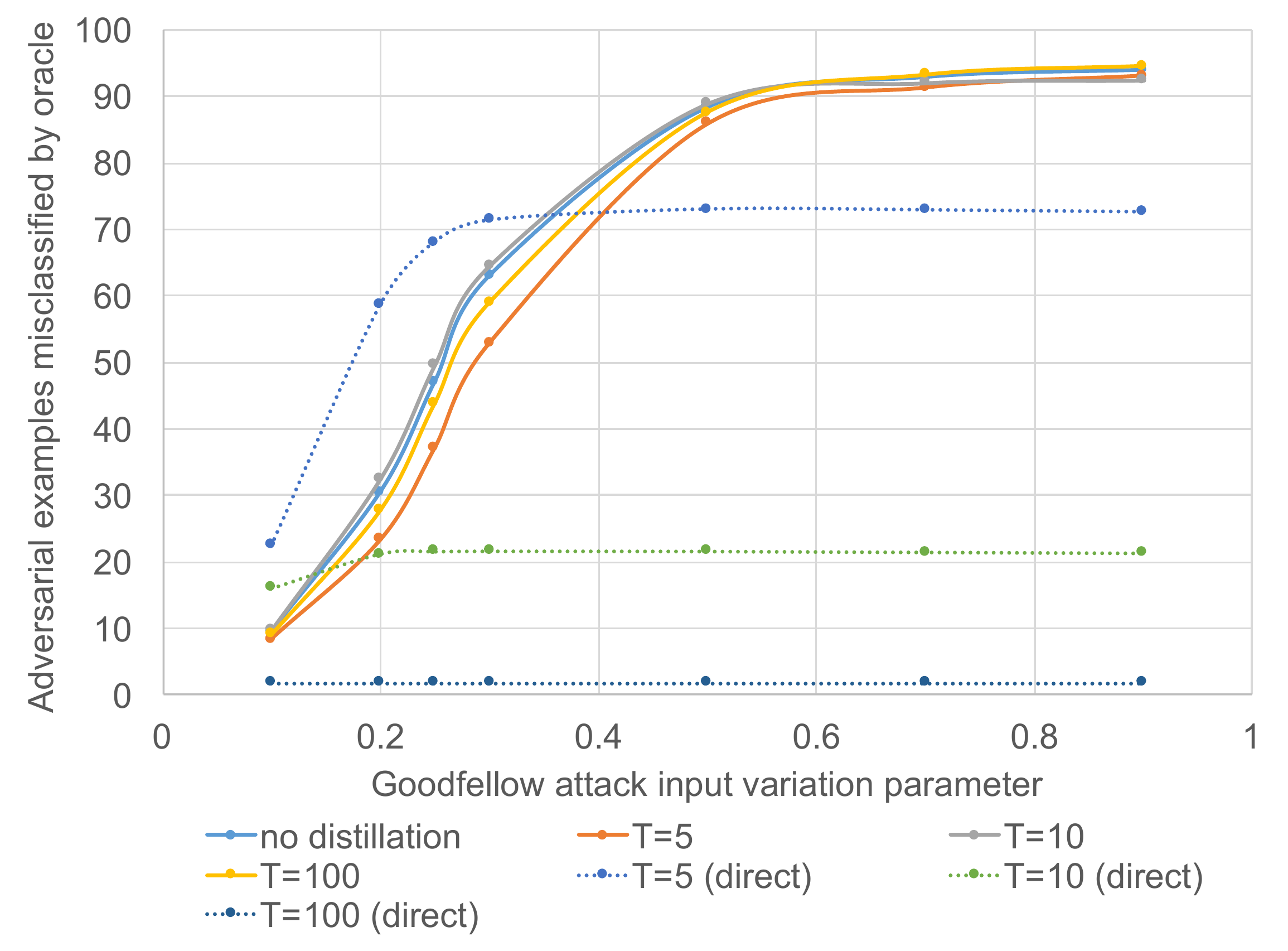}
	\caption{\textbf{Evaluation of defensive distillation:} Percentage of adversarial examples crafted using the Goodfellow algorithm at varying $\varepsilon$ misclassified by the oracle. $T$ is the temperature of distillation~\cite{papernot2015distillation}. Curves marked by (direct) indicate baseline attacks computed on the oracle, all other curves where computed using a substitute, as described in Section~\ref{sec:methodology}. Despite distillation  preventing the attack on the oracle directly, using a substitute allows us to evade it.}
	\label{fig:defensive-distillation-eval}
	\vspace*{-0.1in}
\end{figure}

\boldpara{Defensive distillation} 
Due to space constraints, we refer readers to \cite{papernot2015distillation} for
a detailed presentation of defensive
distillation, which is an alternative defense.
Because the remotely hosted APIs we study here do not implement defensive distillation or provide
primitives that could be used to implement it,
we are forced to evaluate this defense on a locally trained oracle.
Therefore, we train a distilled model as described in~\cite{papernot2015distillation} to act as our MNIST oracle.

We train several variants of the DNN architecture A at different
distillation temperatures $T=5,10,100$.
For each of them, we measure the success of the fast gradient sign attack
(i.e., the Goodfellow et al. algorithm) directly performed on the distilled oracle---as a baseline corresponding to a
white-box attack---and using a substitute DNN trained with synthetic data as
described throughout the present paper.
The results are reported in Figure~\ref{fig:defensive-distillation-eval} for different values of the input variation parameter $\varepsilon$ on the horizontal axis. We find that defensive distillation defends against the fast gradient sign method when the attack is performed directly on the distilled model, i.e. in \emph{white-box settings}. However, in \emph{black-box} settings using the attack introduced in the present paper, the fast gradient sign method is found to be successful regardless of the distillation temperature used by the oracle. We hypothesize that this is due to the way distillation defends against the attack: it reduces the gradients in local neighborhoods of training points.  However, our substitute model is not distilled, and as such possesses the gradients required for the fast gradient sign method to be successful when computing adversarial examples.


Defenses which make models robust in a small neighborhood of the training manifold perform \emph{gradient masking}: they smooth the decision surface  and reduce  gradients used by adversarial crafting in small neighborhoods. However, using a substitute and our black-box approach evades these defenses, as the substitute model is not trained to be robust to the said small perturbations. \emph{We conclude that defending against finite perturbations is a more promising avenue for future work than defending against infinitesimal perturbations. }


\section{Conclusions}
\label{sec:conclusion}

We introduced an attack, based on a novel substitute training algorithm using synthetic data generation, to 
craft adversarial examples misclassified by black-box DNNs. Our work is a 
significant step towards relaxing strong assumptions about adversarial
capabilities made by previous attacks.
We assumed only that the adversary is capable of observing labels
assigned by the model to inputs of its choice.
We validated our attack design by
targeting a remote DNN served by MetaMind, forcing it to misclassify
$84.24\%$ of our adversarial samples. We also conducted an extensive
calibration of our algorithm and generalized it to other ML models by instantiating it against classifiers hosted by Amazon and Google, with success rates of $96.19\%$ and $88.94\%$. Our attack evades a category of defenses, which we call \emph{gradient masking}, previously proposed to increase resilience to adversarial examples. Finally, we provided an intuition for 
adversarial sample transferability across DNNs in Appendix B.


\section{Acknowledgments} 

Nicolas Papernot is supported
by a Google PhD Fellowship in Security.
Research was also supported in part by the Army Research Laboratory,
under Cooperative Agreement Number W911NF-13-2-0045 (ARL Cyber Security
CRA), and the Army Research Office under grant W911NF-13-1-0421.
The views and conclusions contained in this document are those of the
authors and should not be interpreted as representing the official policies,
either expressed or implied, of the Army Research Laboratory or the U.S.
Government. The U.S.\ Government is authorized to reproduce and distribute
reprints for government purposes notwithstanding any copyright notation hereon.


\section*{A. DNN architectures}

\label{sec:appendix-architectures}

Figure~\ref{table:dnn-architectures} provides the specific
DNN architectures used throughout Sections~\ref{sec:validation},~\ref{sec:evaluation}, and~\ref{sec:transferability}.
The first column is the identifier used in the paper to refer
to the architecture. The second and third columns respectively
indicate the input and output dimensionality of the model. Finally, 
each additional column corresponds to a layer of the neural network.

\begin{figure}[h!]
\centering
\begin{footnotesize}
\begin{tabular}{|c|c|c|c|c|c|c|c|c|}

\hline
ID & In & Out  & CM & CM & RL & RL & RL  & S \\ \hline \hline
A & $784$ & $10$ & $32$ &  $64$ &  $200$ & $200$ & - &   $10$  \\ \hline 
B & $3072$ & $43$ & $64$ &  $128$ &  $256$ & $256$ &   - & $43$  \\ \hline 
C & $3072$ & $43$ & $32$ &  $64$ &  $200$ & $200$ &   - & $43$  \\ \hline
D & $3072$ & $43$ & $32$ &  $64$ &  $200$ & $200$ &   - & $43$  \\ \hline
E & $3072$ & $43$ & $64$ &  $64$ &  $200$ & $200$ & $100$ &   $43$  \\ \hline
F & $784$ & $10$ & $32$ &  $64$ &  $200$ &   - & - & $10$  \\ \hline 
G & $784$ & $10$ & $32$ &  $64$ &    - & - & - & $10$  \\ \hline 
H & $784$ & $10$ & $32$ & - &  $200$ & $200$ &   - & $10$  \\ \hline 
I & $784$ & $10$ &  - & - & $200$ & $200$ & $200$ &   $10$  \\ \hline 
J & $784$ & $10$ &  - & - & $1000$ & $200$ &   - & $10$  \\ \hline 
K & $784$ & $10$ &  - & - & $1000$ & $500$ & $200$ &   $10$  \\ \hline 
L & $784$ & $10$ & $32$ &  - & $1000$ & $200$ &  - & $10$  \\ \hline 
M & $784$ & $10$ & $32$ &  - & - & 200s & 200s & $10$  \\ \hline

\end{tabular}\end{footnotesize}
\caption{\textbf{DNN architectures:} ID: reference used in the paper, In: input dimension, Out: output dimension, CM: convolutional layer with 2x2 kernels followed by max-pooling with kernel 2x2, RL: rectified linear layer except for $200s$ where sigmoid units are used, S: softmax layer.}
\label{table:dnn-architectures}
\end{figure}


\section*{B. Intuition behind Transferability}
\label{sec:transferability}

Previous work started explaining why adversarial samples transfer between different 
architectures~\cite{goodfellow2014explaining,szegedy2013intriguing}. Here, we
build an intuition behind transferability based on {\it statistical
hypothesis testing}~\cite{hypothesis-testing} and an analysis of DNN cost gradient sign matrices. A formal treatment is left as
future work.

Recall the perturbation 
in the Goodfellow algorithm. 
Inspecting Equation~\ref{eq:goodfellow-perturbation}, it is clear that, given a sample
$\vec{x}$, the noise added would be the same for two DNNs $F$
and $G$ if $\sgn(\nabla_{\vec{x}}cost(F, \vec{x}, y))$ and
$\sgn(\nabla_{\vec{x}}cost(G, \vec{x}, y))$ were equal. 
These matrices have entries in $\{+1,-1\}$. 
Let us write the space of these matrices as ${\mbox Sgn}_{n \times m}$. 
Assume that the samples $\vec{x}$ are
generated from a population distribution ${\cal D}$ (e.g., in our case 
the distribution from which the images of digits are drawn). The formula $\sgn(\nabla_{\vec{x}}cost(F, \vec{x}, y))$
and ${\cal D}$ induce a distribution ${\cal D}_F$ over ${\mbox Sgn}_{n
  \times m}$ (i.e. randomly draw a sample from the distribution ${\cal
  D}$ and compute the quantity). 
Similarly,  DNN $G$ and 
distribution ${\cal D}$ induce a distribution ${\cal D}_G$ over 
${\mbox Sgn}_{n \times m}$. Our main conjecture is: 
\begin{quote}
For two ``similar'' architectures $F$ and $G$ distributions ${\cal D}_F$
and ${\cal D}_G$ induced by a population distribution ${\cal D}$ are highly correlated.
\end{quote} 
If distributions ${\cal D}_F$ and ${\cal D}_G$ were
independent, then the noise they add during
adversarial sample crafting are independent. In this case, our intuition is
that adversarial samples would not transfer (in the
two cases you are adding noise that are independent). The question is: how to verify our conjecture despite the population distribution ${\cal D}$ being unknown? 

We turn to
statistical hypothesis testing. We can empirically estimate
the distributions ${\cal D}_F$ and ${\cal D}_G$ based on known
samples. First, we generate two sequences of sign matrices $\sigma_1 = \langle
M_1,M_2,\cdots \rangle$ and $\sigma_2 = \langle N_1, N_2, \cdots
\rangle$ using the sample set (e.g. MNIST) for a substitute DNN
$F$ and oracle $G$. Next we pose the following {\it null hypothesis}:
\begin{quote}
$H_N$: The sequences $\sigma_1$ and $\sigma_2$ are drawn from independent
distributions.
\end{quote}
We use standard tests from the statistical hypothesis testing literature to
test the hypothesis $H_N$. If the hypothesis $H_N$ is {\it rejected}, then we
know that the sign matrices corresponding to the two architectures $F$ and $G$
are correlated. 

We describe the test we use. There are several algorithms 
for hypothesis testing: we picked a simple one based on a chi-square
test.  An investigation of other hypothesis-testing 
techniques is left as future work. Let $p_{i,j}$ and $q_{i,j}$ 
be the frequency of $+1$ in the $(i,j)$-th entry of matrices in sequences
$\sigma_1$ and $\sigma_2$, respectively. Let $r_{i,j}$ be the frequency
of the $(i,j)$-th entry being $+1$ in both sequences $\sigma_1$ and $\sigma_2$
 simultaneosuly.\footnote{We assume that the frequencies are normalized so they can
be interprested as probabilities, and also assume that all frequencies are
$>0$ to avoid division by zero, which can be achieved by rescaling.} Note that if the distributions were independent
then $r_{i,j} = p_{i,j} q_{i,j}$. However, if the distributions are correlated, then we expect $r_{i,j} \not= p_{i,j} q_{i,j}$.
Consider quantity:
\[
\chi^{2 \star} = \sum_{i=1}^m \sum_{j=1}^n \frac{ (r_{i,j} N -p_{i,j} q_{i,j} N)^2 }{ p_{i,j} q_{i,j} N}
\]
where $N$ is the number of samples. In the $\chi$-square test, we compute the probability that $P(\chi^2 > \chi^{2 \star})$, where
$\chi^2$ has degrees of freedom $(m-1)(n-1)=27\times 27=729$ for the MNIST data. The $\chi^{2
\star}$ scores for substitute DNNs from Table~\ref{table:goodfellow-transferability-architectures}
range between $61,403$ for DNN A and $88,813$ for DNN G.
Corresponding P-values are below $10^{-5}$ for all architectures, with confidence $p<0.01$.
Thus, for all substitute DNNs, the hypothesis $H_N$ is largely rejected:
sequences $\sigma_1$ ans $\sigma_2$, and therefore sign matrices corresponding
to pairs of a substitute DNN and the oracle, are highly correlated. As a
baseline comparison, we generate $2$ random sign matrices and compute the corresponding $\chi^{2*}$ score:
$596$. We find a P-Value of $0.99$ with a confidence of $0.01$, meaning that
these matrices were indeed drawn from independent distribution.

\begin{figure}[t]
	\includegraphics[width=\columnwidth]{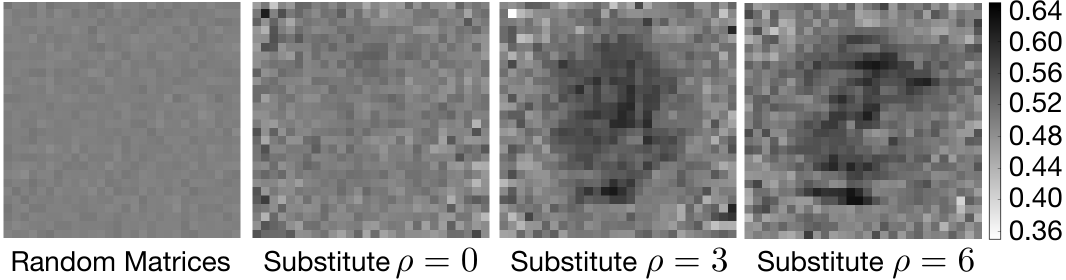}
	\caption{\textbf{Frequencies of cost gradient sign matrix components equal between substitute A and the oracle} at substitute training epochs $\rho\in\{0,3,6\}$ (three on the right), compared to a pair of random sign matrices (first image).}
	\label{fig:sign-correlations}
\end{figure}

\begin{figure}[t]
	\includegraphics[width=\columnwidth]{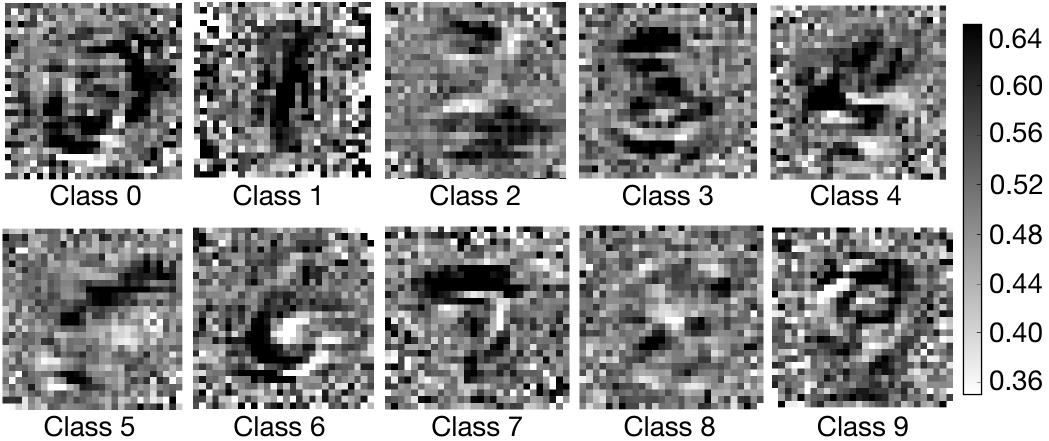}
	\caption{\textbf{Frequencies of cost gradient sign matrix components equal between substitute A and the oracle}}
	\label{fig:sign-correlations-class}
\end{figure}

However, we must now complete our analysis to characterize the correlation
suggested by the hypothesis testing. In Figure~\ref{fig:sign-correlations}, we
plot the frequency matrix $R=[r_{i,j}]$ for several pairs of matrices. The first is a
pair of random matrices of $\{+1,-1\}$. The other matrices correspond to 
substitute DNN A and the oracle at different substitute training epochs $\rho$. Frequencies are computed using the $10,000$ samples of the MNIST test set. Although
all frequencies in the random pairs are very close to $1/2$, frequencies
corresponding to pixels located in the center of the image are higher in the
$(substitute, oracle)$ matrix pairs. The phenomenon amplifies as
 training progresses through the substitute epochs. We then compute the 
 frequencies separately for each sample source
class in Figure~\ref{fig:sign-correlations-class}. Sign matrices agree on pixels relevant for classification in each
class. 
We plotted similar figures for other substitute DNNs. They are not included due to space constraints. 
They show that substitutes yielding lower transferability 
also have less components of their cost gradient 
 sign matrix frequently equal to the oracle's. This suggests that
  \emph{correlations between the respective sign matrices of the substitute DNN
 and of the oracle---for input components that are relevant
 to classification in each respective class---could explain cross-model adversarial sample transferability.}

\vfill\eject

\section*{C. Discussion of Related Work}
\label{sec:discussion}

Evasion attacks against classifiers were discussed previously. Here, we cover below black-box attacks in more details.

Xu et al. applied a genetic algorithm to evade malware detection~\cite{xuautomatically}. 
Unlike ours, it accesses probabilities assigned by the classifier to compute genetic variants fitness. These can be concealed by defenders. The attack is also not very efficient: 500 evading variants are found in 6 days. As the classifier is queried heavily, the authors conclude that the attack cannot be used against remote targets.
Finally, given the attack's high cost on low-dimensional random forests and SVMs, it is unlikely the approach would scale to DNNs. 

Srndic et al. explored the strategy of training a substitute model to find evading inputs~\cite{laskov2014practical}.
They do so using labeled data, which is expensive to collect,
especially for models like DNNs.
In fact, their attack is evaluated only on random forests and an SVM.
Furthermore, they exploit a semantic gap between the specific
classifiers studied and PDF renderers, which prevents their attack from being applicable to models that do not create such a semantic gap.
Finally, they assume knowledge of hand-engineered high-level features whereas
we perform attacks on raw inputs. 

Tramer et al. considered an adversarial goal different from ours: the one of
extracting the exact value of each model parameter. Using partial knowledge of
models and equation solving, they demonstrated how an adversary may recover
parameters from classifiers hosted by BigML and
Amazon~\cite{tramer2016stealing}. However, it would be difficult to scale up the approach 
to DNNs in practice. To recover the $2,225$ parameters of a shallow neural network (one
hidden layer with $20$ neurons) trained on a local machine, they make $108,200$
label queries. Instead, we make $2,000$ label queries to train substitute DNNs
made up of $8$ hidden layers (each with hundreds of neurons) with a total of
over $100,000$ parameters---albeit at the expense of a reduced guaranteed
accuracy for the model extraction operation. Unlike theirs, our work also shows
that our substitutes enable the adversary to craft adversarial examples that are
likely to mislead the remote classifier.

\end{document}